\documentclass[12pt, journal,onecolumn,draftclsnofoot]{IEEEtran}
\usepackage[noadjust]{cite}
\usepackage[cmex10]{amsmath}
\usepackage{amsfonts}
\usepackage{amssymb}
\usepackage{algorithm}
\usepackage{algorithmic}
\usepackage[dvipsnames]{xcolor}
\usepackage[colorlinks,linkcolor=black,anchorcolor=black,citecolor=black]{hyperref}
\usepackage{graphicx}
\usepackage{verbatim}
\usepackage{indentfirst}
\usepackage{epsfig}
\usepackage{bm}
\usepackage{multirow}
\usepackage{subfigure}

\IEEEoverridecommandlockouts
\newtheorem{remark}{\underline{Remark}}[section]
\newtheorem{proposition}{\underline{Proposition}}[section]

\newcommand{\mv}[1]{\mbox{\boldmath{$ #1 $}}}

\begin{document}
\title{Empowering Base Stations with Co-Site Intelligent Reflecting Surfaces: User Association, Channel Estimation and Reflection Optimization}
\author{Yuwei Huang, {\it Student Member, IEEE}, Weidong Mei, {\it Member, IEEE}, and Rui Zhang, {\it Fellow, IEEE}
\thanks{The authors are with the Department of Electrical and Computer Engineering, National University of Singapore, Singapore 117583 (e-mails: yuweihuang@u.nus.edu, wmei@nus.edu.sg, elezhang@nus.edu.sg). Y. Huang is also with the NUS Graduate School, National University of Singapore, Singapore 119077. R. Zhang is also with the School of Science and Engineering, The Chinese University of Hong Kong, Shenzhen 518172, China.}}
\maketitle

\begin{abstract}
Intelligent reflecting surface (IRS) has emerged as a promising technique to enhance wireless communication performance cost-effectively. The existing literature has mainly considered IRS being deployed near user terminals to improve their performance. However, this approach may incur a high cost if IRSs need to be densely deployed in the network to cater to random user locations. To avoid such high deployment cost, in this paper we consider a new IRS aided wireless network architecture, where IRSs are deployed in the vicinity of each base station (BS) to assist in its communications with distributed users regardless of their locations. Besides significantly enhancing IRSs' signal coverage, this scheme helps reduce the IRS-associated channel estimation overhead as compared to conventional user-side IRSs, by exploiting the nearly static BS-IRS channels over short distance. For this scheme, we propose a new two-stage transmission protocol to achieve IRS channel estimation and reflection optimization for uplink data transmission efficiently. In addition, we propose effective methods for solving the user-IRS association problem based on long-term/statistical channel knowledge and the selected user-IRS-BS cascaded channel estimation problem. Finally, all IRSs'  passive reflections are jointly optimized with the BS's multi-antenna receive combining to maximize the minimum achievable rate among all users for data transmission. Numerical results show that the proposed co-site-IRS empowered BS scheme can achieve significant performance gains over the conventional BS without co-site IRS and existing schemes for IRS channel estimation and reflection optimization, thus enabling an appealing low-cost and high-performance BS design for future wireless networks.
\end{abstract}
\section{Introduction}
Intelligent reflecting surface (IRS) is a digitally controllable planar surface consisting of a massive number of passive reflecting elements, whose amplitude and phase shifts can be individually adjusted in real time, thereby enabling a dynamic control over the wireless propagation environment. IRS can thus be applied in wireless communication systems to achieve various new functions, such as channel reconfiguration, passive relaying, passive interference nulling/cancellation, etc \cite{tutorial_new,tutorial,magazine}. Since IRS dispenses with radio frequency (RF) chains and only reflects the ambient signals as a passive array, it features low hardware cost and power consumption, thus deemed as a promising candidate for next-generation wireless communication technologies (e.g., sixth-generation (6G)) \cite{6G,6G2}. Motivated by this, IRS has been extensively investigated in the literature under various system setups and for different design objectives, such as multiple-input multiple-output (MIMO) communications \cite{array,mimo2}, cell-free networks \cite{cell_free,cell_free1}, wireless information and power transfer \cite{swipt1,swipt2}, orthogonal frequency division multiplexing (OFDM) based broadband systems \cite{ofdm,ofdm2}, non-orthogonal multiple access (NOMA) \cite{NOMA1,NOMA2}, multi-cell network \cite{sca,multi_cell}, mobile edge computing \cite{MEC1,MEC2}, physical-layer security \cite{security1,security2}, unmanned aerial vehicle (UAV)-ground communications \cite{UAV1,UAV2}, among others.  

To fully reap the passive reflection gain of IRS, it is crucial to acquire the channel state information (CSI) accurately, which, however, is practically challenging due to the lack of active RF chains at IRS reflecting elements as well as their large number in practice. In the literature, there are two main approaches for IRS channel estimation based on different IRS configurations, namely semi-passive IRS and fully passive IRS \cite{tutorial,magazine}. In the first case, additional sensing devices (e.g., low-power sensors) equipped with low-cost receive RF chains are integrated into IRS. By this means, the user-IRS and base station (BS)-IRS channels can be separately estimated for the uplink and downlink communications based on the pilot signals sent by the users and BS, respectively \cite{huxiaoling}. Although this approach works well in the time-division duplexing (TDD) mode by exploiting the channel reciprocity to estimate the CSI from the IRS to users/BS, it becomes ineffective in the frequency-division duplexing (FDD) mode\footnote{It may be practically possible to estimate the channel from the IRS to the users/BS if active sensors that can both transmit and receive signals are integrated into the IRS, which, however, results in considerably higher hardware cost and power consumption.}. In contrast, for the second case, with the fully passive IRS, although the BS-IRS and user-IRS channels cannot be estimated individually as by the semi-passive IRS, the cascaded user-IRS-BS or BS-IRS-user channel can be estimated at the BS or each user based on the pilot signals sent by the other, which is applicable to both TDD and FDD modes \cite{estimation1,estimation2,estimation3}. However, as compared to the semi-passive IRS, this approach entails substantially more channel parameters to be estimated due to the redundancy in cascaded channels, especially when the number of IRSs, IRS reflecting elements, and/or users becomes large. To reduce the channel training time for the fully passive IRS, assorted methods have been proposed in the literature, including IRS elements grouping \cite{ofdm,group,DFT2}, reference user based channel estimation \cite{liuliang,DFT2}, anchor-aided channel estimation \cite{anchor}, channel estimation based on channel sparsity \cite{chenjie,compressed}, and so on (see \cite{tutorial,tutorial2,tutorial3} and the references therein).

\begin{figure}
\centering
\includegraphics[width=10cm]{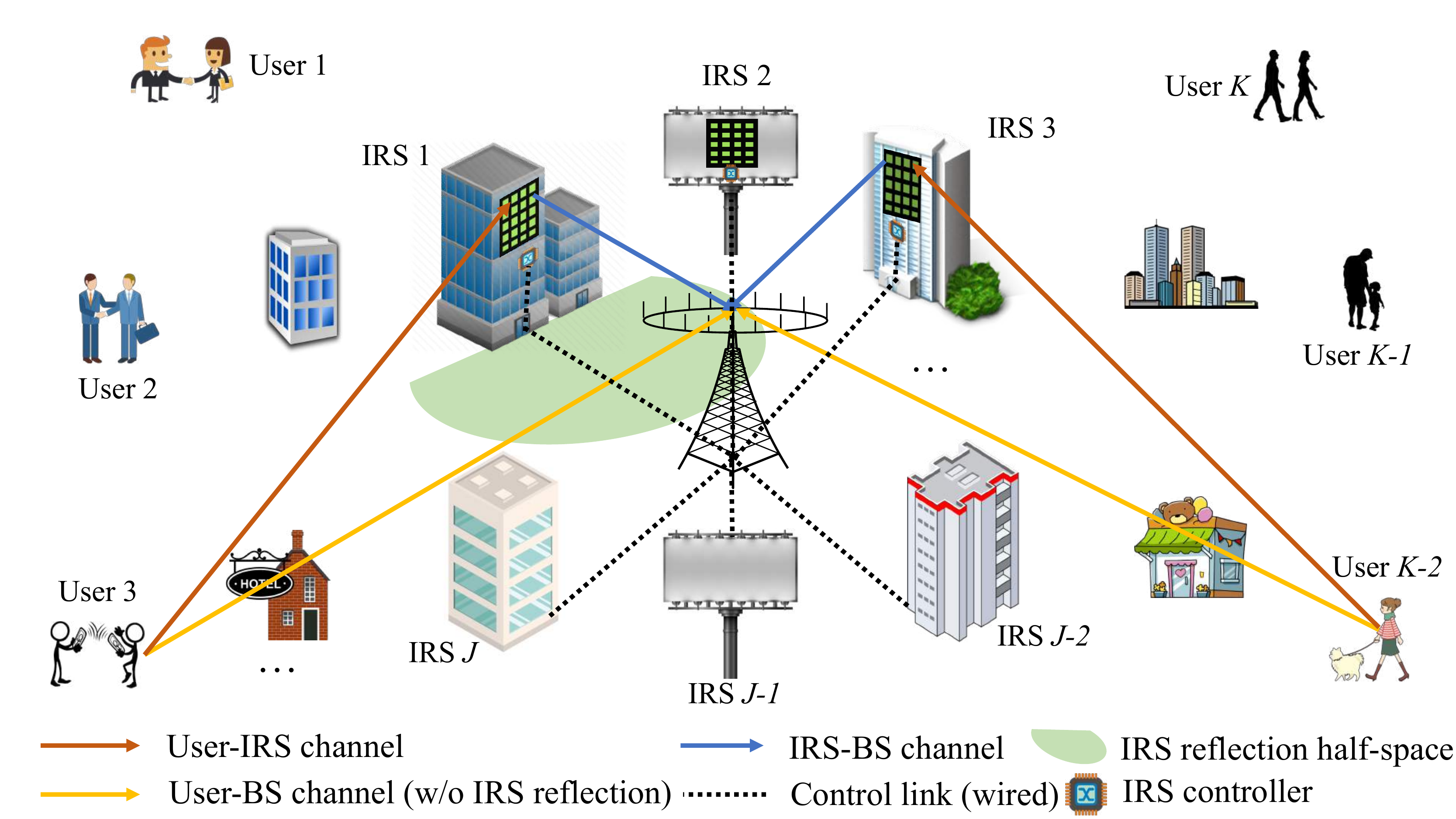}
\caption{A new architecture of BS empowered by multiple co-site IRSs in its vicinity.}\label{system_model}
\vspace{-9pt}
\end{figure}
On the other hand, the deployment of IRSs also has a significant effect on the communication performance. Existing works have mostly considered IRSs being deployed closely to user terminals (e.g., at hotspot, cell edge, and in moving vehicles) for improving their communication rates. However, to cater to the random user locations, IRSs need to be densely deployed in the network, which inevitably incurs an exorbitant deployment cost and signaling overhead between the BS and the IRSs' controllers. In addition, due to the IRS-induced double path loss, the rate performance of each user mainly depends on the passive reflection gain from its nearest IRS, which may be insufficient if their channel is severely blocked. To overcome the above limitations of user-side IRSs, in this paper, we consider a new {\it co-site-IRS empowered BS} architecture, where multiple distributed IRSs are deployed in the vicinity of each BS and assist in its communications with different users, as shown in Fig. \ref{system_model}. As compared to user-side IRSs, co-site IRSs with the BS enjoy the following main advantages \cite{deploy,deploy2}. First, as each BS-side IRS can effectively reflect signals to/from half of the space (see Fig. \ref{system_model}), each user can be potentially served by much larger number of IRSs at any time, thus significantly enhancing IRSs' reflected signal coverage and increasing their passive  reflection gain achievable for each user. Second, the short BS-IRS distances greatly reduce the control link distances between the BS and IRS controllers and thus the signaling overhead among them. Third, due to the fixed locations of the BS and all IRSs, which are in short distance and also at high altitude above the ground (see Fig. \ref{system_model}), the BS-IRS channels are typically line-of-sight (LoS) dominant and remain static for a very long period, which is more reliable as compared to user-side IRSs and can also be leveraged to reduce the training overhead for user-IRS-BS cascaded channel estimation in real time (as will be shown in this paper).

Furthermore, as compared to the conventional multi-antenna BS without co-site IRS, the co-site-IRS empowered BS may offer the same user performance, but with greatly reduced number of active antennas at the BS, by exploiting the passive reflection and spatial multiplexing gains provided by distributed nearby IRSs. This thus helps decrease the hardware cost and energy consumption of multi-antenna BSs, which is known as a major obstacle to their ultra-large array implementation in future wireless systems, especially for those operating at high frequencies such as millimeter wave (mmWave) frequency. In order to enable the proposed co-site-IRS empowered BS architecture, in this paper, we propose a new two-stage transmission protocol for it to achieve efficient IRS channel estimation and reflection optimization for data transmission in the uplink. The main contributions of this paper are summarized as follows.
\begin{itemize}
\item First, for the first stage of the proposed protocol, we introduce a new cooperative channel training scheme by invoking all IRSs' controllers to acquire useful long-term channel knowledge of the cascaded IRS controller-IRS-BS channels (which can be considered as additional user-IRS-BS channels by treating each IRS controller as an equivalent user) as well as the statistical CSI on all user links. The former will be utilized as reference CSI to reduce the overhead of real-time channel estimation for the users in the second stage. While the latter is utilized to determine the set of users served by each IRS, based on which the set of user-IRS-BS cascaded channels that need to be estimated in real time in the second stage are optimized, subject to a given total channel training duration for all users. This problem is referred to as user-IRS association, for which we propose a large-scale performance metric based solution by maximizing the minimum performance metric among all users via a successive convex approximation (SCA) algorithm.
\item Next, for the second stage of the proposed protocol, we devise a new method to estimate user-IRS-BS cascaded channels efficiently based on the designed user-IRS associations, where the IRS controller-IRS-BS reference CSI obtained in the first stage is used to shorten the training time. Based on the estimated CSI, all IRSs' passive reflections are jointly optimized to maximize the minimum average signal-to-interference-plus-noise-ratio (SINR) among all users, by accounting for the CSI which is not estimated for those cascaded user-IRS-BS channels of non-associated user-IRS pairs. Although this problem is challenging to solve, we obtain a locally optimal solution by applying the alternating optimization (AO) and gradient projection method (GPM).
\item Last, numerical results are provided, which show that the proposed co-site-IRS empowered BS design yields considerably higher achievable rate as compared to the conventional multi-antenna BS without co-site IRS, as well as the existing schemes for IRS channel estimation and reflection optimization. The effects of key system parameters, such as the channel coherence time, channel training duration, and total number of IRS reflecting elements, are also investigated and discussed. The results show that the proposed co-site-IRS empowered BS can be a compelling candidate for achieving low-cost and high-performance wireless networks in the future.
\end{itemize}

It is worth noting that there have been several recent works on studying the multi-IRS aided wireless network, where multiple IRSs are employed to enhance the transmission of each wireless link by leveraging their successive signal reflections and cooperative passive beamforming (see e.g., \cite{multiple1,multiple2,multiple3,multiple4,multiple5}). Different from this line of work, in this paper, we consider one dominant signal reflection by each IRS only and ignore their inter-reflections by deploying the IRSs sufficiently far apart from each other. The results of this paper can be extended to the general case with inter-IRS signal reflection in future work. It should also be mentioned that there exist some prior works focusing on the joint channel estimation and IRS reflection optimization \cite{estimation2, joint_estimation_reflection,joint_estimation_reflection2}. However, these works only considered the scenario with a single or multiple {\it user-side} IRSs and did not leverage the unique features of the proposed IRS-empowered BS architecture to improve the user rate performance with reduced channel estimation overhead.

The rest of this paper is organized as follows. Section II presents the system model and the proposed two-stage transmission protocol for the co-site-IRS empowered BS. Section III presents the long-term channel estimation and user-IRS association optimization in the first stage of the proposed protocol. Section IV presents the short-term cascaded user-IRS-BS channel estimation and IRS passive reflection optimization in the second stage of the proposed protocol. Section V presents numerical results to verify the efficacy of our proposed design. Finally, Section VII concludes this paper.

{\it Notation:} In this paper, scalars, vectors and matrices are denoted by italic, bold-face lower-case and bold-face upper-case letters, respectively. The transpose and conjugate transpose of a matrix are denoted as $(\cdot)^{T}$ and $(\cdot)^{H}$, respectively. $(\mv a)_{m}$ denotes the $m$-th entry of a vector $\mv a$, and $[\mv G]_{m,n}$ denotes the entry at the $m$-th row and $n$-th column of a matrix $\mv G$. $\mv I_{M}$ denotes the identity matrix of size $M$. $\mathbb{C}^{x\times y}$ denotes the set of $x$-by-$y$ complex-valued matrices. For a complex number $s$, $s^{*}$, $|s|$, and $\angle{s}$ denote its conjugate, amplitude and phase, respectively, and $s\sim\mathcal{CN}(0,\sigma^{2})$ means that it is a circularly symmetric complex Gaussian (CSCG) random variable with zero mean and variance $\sigma^{2}$. $\mathbb{E}[\cdot]$ denotes the statistical expectation. $|\mathcal {A}|$ denotes the cardinality of a set $\mathcal {A}$. $i$ denotes the imaginary unit, i.e., $i^2=-1$. For a vector $\mv a$, $\text{diag}(\mv a)$ denotes a diagonal matrix whose diagonal elements are specified by $\mv a$, and $\lVert \mv a \rVert$ denotes its norm. The notation $\log_{2}(\cdot)$ denotes the logarithm function with base $2$. $\mathcal{O}(\cdot)$ denotes the Landau's symbol to describe the order of complexity.

\section{System Model and Transmission Protocol}\label{system_model_section}
\subsection{System Model}
Fig. \ref{system_model} depicts the architecture of the proposed co-site-IRS empowered BS in a wireless network, where the BS serves $K$ single-antenna users with the help of $J$ IRSs deployed in its vicinity. In this paper, we consider the uplink communications from the users to the BS, while the results of this paper can be also applied to the downlink communications if TDD is used for separating the uplink and downlink communications where the uplink-downlink channel reciprocity holds for all the links. In practice, as an IRS can only reflect the signals incident on its front reflection half-space, all IRSs are deployed to face the BS, so as to effectively reflect the signals to/from it. However, due to the random locations of the users, each IRS usually can only reflect the signals from/to the users located in its front reflection half-space effectively\footnote{Note that if a user is located outside the reflection half-space of any IRS, its transmitted signal may be first randomly scattered in the environment before reaching the reflection surface of this IRS, which will result in much higher path loss in practice.}. Generally, increasing the number of IRSs, $J$, can help improve the system performance due to the enhanced aperture and/or reflected signal coverage. Furthermore, we consider that a smart controller is attached to each IRS, which is able to exchange information instantly with the BS via a high-speed wired or wireless link with negligible delay thanks to their proximity. Each IRS controller is assumed to be placed sufficiently close to its controlled reflecting elements (see Fig. \ref{system_model}), such that its large-scale channel gain with the BS/users is approximately equal to that of its controlled reflecting elements. The BS and each IRS are assumed to be equipped with $M$ antennas and $N$ reflecting elements, respectively. For convenience, we denote the sets of IRSs (or IRS controllers), users, BS antennas, and reflecting elements of each IRS as $\mathcal {J}\triangleq \{1,2,\cdots,J\}$, $\mathcal {K}\triangleq\{1,2,\cdots,K\}$, $\mathcal {M}\triangleq\{1,2,\cdots,M\}$, and $\mathcal {N}\triangleq\{1,2,\cdots,N\}$, respectively.

In this paper, we consider quasi-static block-fading channels and focus on a given fading block, during which all the channels involved are assumed to be constant. Let $\mv h_{d,k}=[h_{d,k,1},\cdots,h_{d,k,M}]^{T}\in\mathbb{C}^{M\times 1}$ denote the channel from user $k$ to the BS in the considered block, $\mv t_{k,j}=[t_{k,j,1},\cdots,t_{k,j,N}]^{T}\in\mathbb{C}^{N\times 1}$ denote that from user $k$ to IRS $j$, and $\mv F_{j}=[\mv f_{j,1},\cdots,\mv f_{j,N}]\in\mathbb{C}^{M\times N}$ denote that from IRS $j$ to the BS. Their distance-dependent large-scale channel gains are denoted as $\beta_{k}^{2}$, $\alpha_{k,j}^{2}$, and $\mu_{j}^{2}$, respectively. Note that thanks to the proximity of the BS and IRS as well as their fixed locations, the IRS-BS channels, i.e., $\mv F_{j},~j\in\mathcal J$, generally vary much slower with time and enable more favorable LoS propagation as compared to the user-associated channels, i.e., $\mv h_{d,k}$ and $\mv t_{k,j},~k\in\mathcal K,j\in\mathcal J$. This fact will be exploited later in this paper to reduce the overhead of real-time channel estimation in our proposed transmission protocol. Let $\mv\Theta_{j}=\text{diag}(\mv\theta_{j})$ denote the reflection matrix of IRS $j$, $j\in\mathcal J$, where $\mv \theta_{j}=[e^{i\theta_{j,1}},\cdots,e^{i\theta_{j,N}}]^{T}$ denotes its reflection vector and $\theta_{j,n}\in[0,2\pi)$ denotes the phase shift by its $n$-th reflecting element. Note that here we have set the amplitude of all reflection coefficients to its maximum value of one for the purpose of maximizing reflected signal power\footnote{An additional control of the amplitude of each reflecting element may further improve the system performance by providing more design flexibility \cite{amplitude}. However, this requires more sophisticated hardware designs and thus higher control complexity \cite{tutorial}.}. Hence, the overall channel from user $k$ to the BS is given by\footnote{Notice that the multi-hop IRS reflections may also exist in the overall user-BS channels. However, their effect would be marginal under the proposed designs based only on IRSs' single-hop reflections. This is because no passive beamforming gain can be achieved via their multi-hop reflections in this case to compensate more severe multiplicative path loss.}
\begin{align}
\mv h_{k}(\{\mv\theta_{j}\})=\mv h_{d,k}+\sum_{j=1}^{J}\mv F_{j}\mv\Theta_{j}\mv t_{k,j}=\mv h_{d,k}+\sum_{j=1}^{J}\mv G_{k,j}\mv\theta_{j},\label{effective}
\end{align}
where $\mv G_{k,j}\triangleq\mv F_{j}\text{diag}(\mv t_{k,j})=[\mv g_{k,j,1},\cdots,\mv g_{k,j,N}]\in\mathbb{C}^{M\times N}$ denotes the cascaded channel from user $k$ to the BS via IRS $j$, with $\mv g_{k,j,n}=\mv f_{j,n}t_{k,j,n}$ denoting its $n$-th column.

It is noted from (\ref{effective}) that the design of all IRSs' passive reflections, i.e., $\mv\theta_{j},~j\in\mathcal J$, in general requires the knowledge of all direct BS-user channels $\mv h_{d,k},~k\in\mathcal K$, and all cascaded user-IRS-BS channels $\mv G_{k,j},~k\in\mathcal K,j\in\mathcal J$, which involves $KM+KMNJ$ unknown parameters in total. As such, this may incur excessively high overhead for channel estimation if $K$, $N$ and/or $J$ is practically large, which will greatly reduce the time for data transmission due to the finite duration of each fading block. Thus, efficient channel estimation is crucial to the proposed co-site IRSs for enhancing the BS's capacity.

\subsection{Transmission Protocol}
To implement the co-site-IRS empowered BS efficiently, we propose a new transmission protocol to reduce the overhead of cascaded channel estimation and facilitate the IRS reflection design, based on the following two important features of BS-side IRSs. 

{\it First,} some cascaded channels may only have a marginal effect on a given user's rate performance. On one hand, if user $k$ is located outside the reflection half-space of IRS $j$, then $\mv G_{k,j}$ usually has a much weaker strength than $\mv G_{k,q}$'s for IRS $q$'s $q\neq j$ which face user $k$ in their reflection half spaces. Hence, $\mv G_{k,j}$ may not need to be estimated without affecting the achievable rate of this user. On the other hand, to balance the rate performance of all users, the IRS passive reflection should cater to the users having weak direct channels with the BS with higher priority as compared to those with stronger direct channels with the BS, which implies that the latter's cascaded channels may not need to be fully estimated or even estimated at all. Motivated by the above, instead of estimating all cascaded channels for all users with all IRSs, we only need to select a subset of users to be served by each IRS and only their cascaded channels need to be estimated. For convenience, if the cascaded user $k$-IRS $j$-BS channel, i.e., $\mv G_{k,j}$, is estimated in our protocol, we consider that user $k$ can be associated with IRS $j$ as its reflection design will be based on the estimated $\mv G_{k,j}$. Accordingly, to describe the user-IRS associations, we define a set of binary variables, i.e., $\delta_{k,j},~k\in\mathcal K,j\in\mathcal J$, for which $\delta_{k,j}=1$ if user $k$ is associated with IRS $j$; otherwise, we set $\delta_{k,j}=0$.

{\it Second}, it is noted that the cascaded channels from different users to the BS via {\it any IRS} share the same IRS-BS channel, despite that their user-IRS channels may be different. For example, the cascaded channels from two users $k$ and $q$ to the BS via the $n$-th element of IRS $j$ are given by $\mv g_{k,j,n}=\mv f_{j,n}t_{k,j,n}$ and $\mv g_{q,j,n}=\mv f_{j,n}t_{k,j,n}$, respectively, which share the same/common IRS-BS channel $\mv f_{j,n}$. In \cite{liuliang}, this fact has been exploited to reduce the overhead of cascaded channel estimation for user-side IRSs. Specifically, suppose that each IRS $j,~j\in\mathcal J$ in our considered system is a user-side IRS. In IRS $j$'s local coverage region, a reference user, denoted as user $U_{j},~U_{j}\in\mathcal K$, is first selected and its cascaded channel with the BS via IRS $j$, i.e., $\mv g_{U_{j},j,n},~n\in\mathcal N$, is estimated. Then, the cascaded channel of each user $k,~k\neq U_{j}$ in the same coverage region of IRS $j$ can be expressed as the scaled version of that of reference user $U_{j}$, i.e., $\mv g_{k,j,n}=\mv f_{j,n}t_{k,j,n}=\mv f_{j,n}t_{U_{j},j,n}t_{k,j,n}/t_{U_{j},j,n}=\mv f_{j,n}t_{U_{j},j,n}\nu_{k,j,n}=\mv g_{U_{j},j,n}\nu_{k,j,n},~n\in\mathcal N$, where $\nu_{k,j,n}\triangleq t_{k,j,n}/t_{U_{j},j,n},~n\in\mathcal N$. As a result, it suffices to estimate the scaling factors, i.e., $\nu_{k,j,n},~n\in\mathcal N$ (scalars), rather than the exact channels $\mv g_{k,j,n},~n\in\mathcal N$ (vectors) for these users, i.e., $k\neq U_{j}$. Similarly, this channel estimation approach can be applied for estimating user-IRS-BS cascaded channels for BS-side IRSs in our context by exploiting the common IRS-BS channels. However, as compared to the common IRS-BS channels in \cite{liuliang} for user-side IRSs, their counterparts for our considered BS-side IRSs are more likely to be static due to much shorter link distance. Moreover, for each IRS $j$, the BS can properly select the controller of a nearby IRS (instead of a mobile user as in \cite{liuliang}) as its ``reference user'' $U_{j}$ and obtain its cascaded CSI, i.e., $\mv g_{U_{j},j,n},~n\in\mathcal N$, which thus leads to more reliable reference CSI estimation (thanks to the much shorter link distance as well as higher transmit power of IRS controller than user terminal in practice). This is crucial to estimate the users' cascaded CSI via this IRS accurately in real time.

\begin{figure}
\centering
\includegraphics[width=10cm]{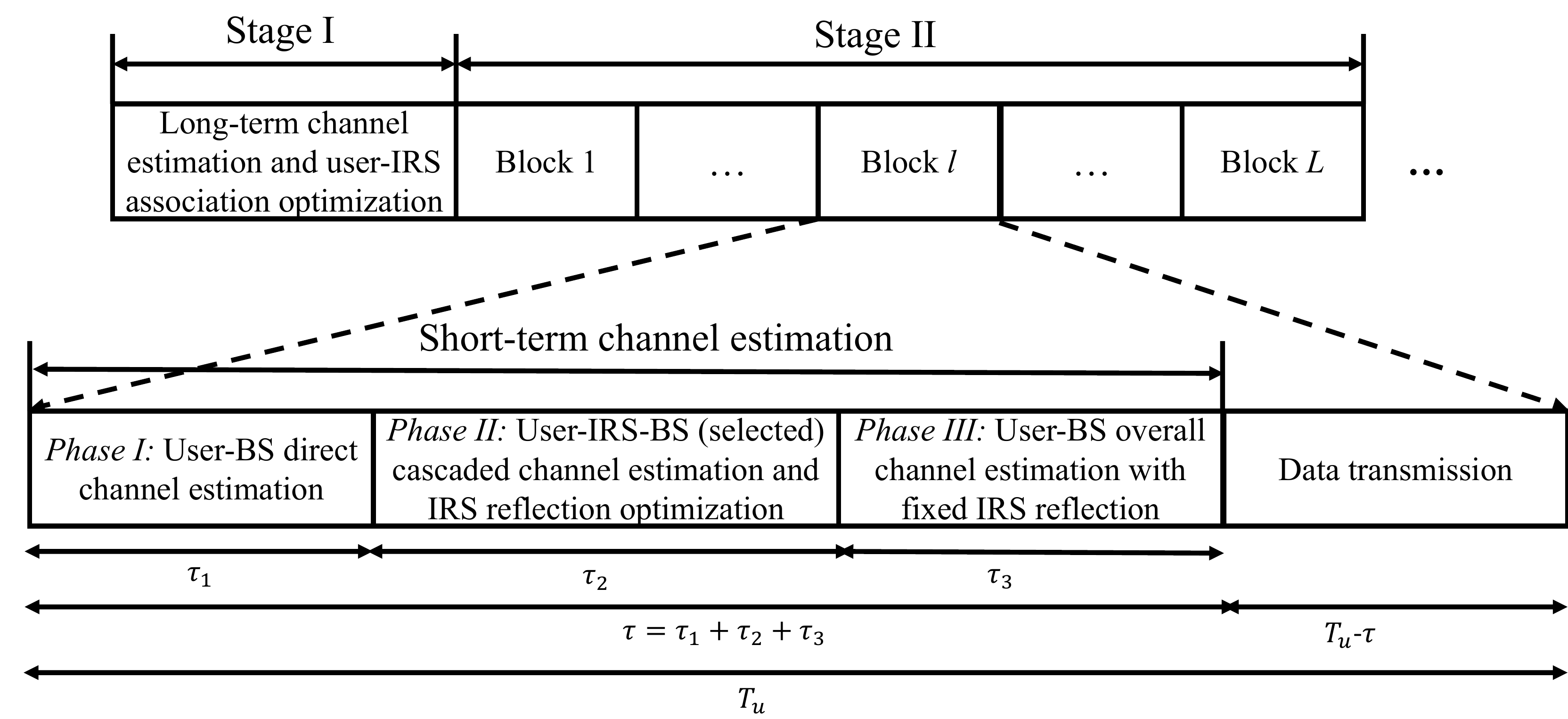}
\caption{Frame structure of the proposed transmission protocol.}\label{protocol}
\end{figure}

Based on the above, we propose a new transmission protocol for each transmission frame in the considered new architecture of co-site-IRS empowered BS, as shown in Fig. \ref{protocol}. The proposed protocol consists of two stages, namely, Stage I and Stage II. In Stage I, the BS cooperates with the IRS controllers to estimate useful long-term channel knowledge, including cascaded IRS controller-IRS-BS channels and statistical CSI on all links, both of which are assumed to remain nearly static during each transmission frame. Accordingly, the former can be utilized as the reference CSI to assist in the real-time cascaded channel estimation for all users in this frame. While the latter can be utilized by the BS to optimize the user-IRS associations, i.e., $\delta_{k,j},~k\in\mathcal K,j\in\mathcal J$ and thereby, determine the cascaded channels that need to be estimated in real time in Stage II. The details of the long-term channel estimation and the user-IRS association optimization will be presented in {\bf Sections \ref{user_IRS}-A and \ref{user_IRS}-B}, respectively. 

In Stage II, short-term channel training and data transmission are conducted over $L$ consecutive fading blocks, as shown in Fig. \ref{protocol}. In each fading block, let $T_{u}$ and $\tau$ (both normalized to the symbol period and assumed to be integers) denote its total duration and the duration for (short-term) channel training, respectively. Hence, the duration for data transmission in each block is $T_{u}-\tau$. In particular, the short-term channel training can be further divided into the following three phases.
\begin{itemize} 
\item{\it Phase I:} First, by turning off all IRSs\footnote{The ON/OFF control of each IRS reflecting element can be achieved by applying a binary 0--1 amplitude control module. In other words, by setting the amplitude to zero/one, the IRS is equivalent to being turned off/on. In practice, this can be realized by adjusting the load resistance/impedance in each element \cite{on_off_control}.}, all users send $\tau_{1}\geq K$ orthogonal pilot symbols to the BS for it to estimate the uplink direct channels (without any IRS reflections) as in the conventional system without IRSs.
\item{\it Phase II:} Next, by sequentially turning on IRSs, the selected cascaded channels $\mv G_{k,j}$'s (specified by the user-IRS association designed in Stage I) are estimated, where the cascaded IRS controller-IRS-BS channels estimated in Stage I will be leveraged to facilitate the channel estimation. This phase incurs an overhead of $\tau_{2}$ symbols, and its details will be presented in {\bf Section \ref{ps_design}-A}. Based on the estimates of the direct channels and selected cascaded channels, the passive reflections of all IRSs are jointly optimized at the BS and then informed to the corresponding IRS controllers, as will be detailed in {\bf Section \ref{ps_design}-B}.
\item {\it Phase III:} Last, each IRS controller tunes the passive reflection of its controlled reflecting elements accordingly. All users send $\tau_{3}\geq K$ orthogonal pilot symbols\footnote{To focus on the more challenging cascaded channel estimation, in this paper we assume that the channel estimates of the direct/overall user-BS channels in Phase I/III are perfect.} to the BS again for it to estimate the overall user-BS channels with fixed IRSs' reflections (see (\ref{effective})).
\end{itemize}
It follows from the above that the total channel training duration is given by $\tau=\tau_{1}+\tau_{2}+\tau_{3}$. Hence, with any given $\tau_{1}$ and $\tau_{3}$, a longer $\tau$ leads to a longer $\tau_{2}$, and thus more cascaded channels can be estimated in Stage II. Let $\zeta$ denote the total number of cascaded user-IRS-BS channels that need to be estimated in Stage II. Its value depends on the specific channel training duration $\tau$ for each fading block in Stage II and will be specified in {\bf Section \ref{ps_design}-A}. 

Finally, based on the estimates of the overall user-BS channels given in (\ref{effective}), the BS computes its receive combining weights for each user for data transmission. Denote by $\mv w_{k}\in\mathbb{C}^{M\times 1}$ its receive combining weights for user $k,~k\in\mathcal K$, which is normalized to satisfy $\lVert \mv w_{k} \rVert^{2}=1$. To maximize the SINR of each user $k,~k\in\mathcal K$, the BS adopts the optimal minimum mean square error (MMSE)-based combining over the received signals, i.e.,
\begin{align}
&\mv w_{k}=\frac{\Big(\sum\limits_{q\neq k}\mv h_{q}(\{\mv\theta_{j}\})\mv h_{q}(\{\mv\theta_{j}\})^{H}+\sigma^{2}\mv I_{M}\Big)^{-1}\mv h_{k}(\{\mv\theta_{j}\})}{\lVert\Big(\sum\limits_{q\neq k}\mv h_{q}(\{\mv\theta_{j}\})\mv h_{q}(\{\mv\theta_{j}\})^{H}+\sigma^{2}\mv I_{M}\Big)^{-1}\mv h_{k}(\{\mv\theta_{j}\})\rVert},~k\in\mathcal K.\label{wk_final}
\end{align}
As a result, the SINR achievable by user $k$ is given by
\begin{align}
&\gamma_{k}(\mv w_{k},\{\mv\theta_{j}\})=\frac{p|\mv w_{k}^{H}\mv h_{k}(\{\mv\theta_{j}\})|^{2}}{\sigma^{2}+p\sum\limits_{q\neq k}|\mv w_{k}^{H}\mv h_{q}(\{\mv\theta_{j}\})|^{2}},\label{sinr_k}
\end{align}
where $p$ denotes the transmit power of each user\footnote{The proposed approaches are also applicable to different user transmit power, by absorbing its effect into the overall channels, $\mv h_{k}(\{\mv\theta_{j}\})$.}, and $\sigma^{2}$ is the noise power at each BS's receiving antenna. Accordingly, the achievable rate of user $k$ in bits/second/Hertz (bps/Hz) is given by
\begin{align}\label{rate}
&R_{k}(\mv w_{k},\{\mv\theta_{j}\})=\frac{T_{u}-\tau}{T_{u}}\log_{2}\Bigg(1+\frac{p|\mv w_{k}^{H}\mv h_{k}(\{\mv\theta_{j}\})|^{2}}{\Gamma(\sigma^{2}+p\sum\nolimits_{q\neq k}|\mv w_{k}^{H}\mv h_{q}(\{\mv\theta_{j}\})|^{2})}\Bigg),
\end{align}
where $\Gamma \ge 1$ characterizes the rate loss due to the practical modulation and coding scheme.

\begin{remark}\label{tradeoff_tau}
Note that when the training duration $\tau$ increases, since more cascaded channels can be selected for estimation in Stage II, the BS/IRS combining/reflection design will be refined, which helps enhance the user SINRs in (\ref{sinr_k}). However, this reduces the data transmission time, i.e., $T_{u}-\tau$, and may degrade the user achievable rates in (\ref{rate}). On the other hand, if $\tau$ decreases, the data transmission time can be prolonged while the user SINRs may be compromised as less cascaded channels are estimated in Stage II. Therefore, there generally exists an inherent tradeoff in selecting $\tau$ to maximize the achievable rates in (\ref{rate}), as will be shown later in Section \ref{simulation_results} by simulation results.
\end{remark}

\section{Long-Term Channel Estimation and User-IRS Association Optimization}\label{user_IRS}
In this section, we first present the procedure of long-term channel estimation in Stage I of the proposed transmission protocol (see Fig. \ref{protocol}), including the estimation of the (nearly) static cascaded channels over the IRS controller-IRS-BS links (to be used for the cascaded channel estimation for the users in Stage II) and the acquisition of the statistical CSI on all involved links. Then, based on the statistical CSI obtained, we propose a large-scale performance metric to determine the user-IRS associations, i.e., $\delta_{k,j},~k\in\mathcal K,j\in\mathcal J$, which also specifies the set of cascaded channels that need to be estimated in Stage II.

\subsection{Long-Term Channel Estimation}
First, to estimate the cascaded channels for the ``reference users'', i.e., $\mv G_{U_{j},j},j\in\mathcal J$, we turn on each of the IRSs sequentially with all the other IRSs turned off (while their controllers are kept on). For example, when IRS $j,~j\in\mathcal J$ is turned on, the BS selects one of its adjacent IRS's controller (which needs to lie in its reflection half-space) as its equivalent reference user $U_{j}$, which then sends pilot symbols to the BS to estimate their cascaded channel via IRS $j$, i.e., $\mv G_{U_{j},j}$. The pilot sequence design of IRS controller $U_{j}$ and reflection design of IRS $j$ in this process can be found in \cite{liuliang} and omitted here for brevity. Let $\hat{\mv G}_{U_{j},j}=[\hat{\mv g}_{U_{j},j,1},\cdots,\hat{\mv g}_{U_{j},j,N}]$ denote the estimate of  $\mv G_{U_{j},j}$, which will be utilized to estimate the users' cascaded channels in Stage II.   

Next, we present how to acquire the statistical CSI which is needed in the user-IRS association optimization, i.e., the large-scale channel gains $\mu_{j}^{2}$, $\alpha_{k,j}^{2}$, and $\beta_{k}^{2}, k\in\mathcal K, j\in\mathcal J$. On one hand, $\mu_{j}^{2},~j\in\mathcal J$ can be measured during the above long-term cascaded channel estimation. Specifically, when IRS controller $U_{j}$ sends its pilot symbols, IRS controller $j$ measures the average received signal strength (RSS) from it, thereby estimating the average channel gain between them, denoted as $\alpha_{U_{j},j}^{2}$. Moreover, note that the cascaded IRS controller $U_{j}$-IRS $j$-BS channel gain is equal to the product of that from IRS controller $U_{j}$ to IRS $j$ (i.e., $\alpha_{U_{j},j}^{2}$) and that from IRS $j$ to the BS (i.e., $\mu_{j}^{2}$); while it can also be computed as the average of all entries of the estimated cascaded channel, i.e., $\hat{\mv G}_{U_{j},j}$, denoted as $g_{U_{j},j}=\sum_{m=1}^{M}\sum_{n=1}^{N}|[\hat{\mv G}_{U_{j},j}]_{m,n}|^{2}/MN$. Hence, we obtain  $\mu_{j}^{2}= g_{U_{j},j}/\alpha_{U_{j},j}^{2},~j\in\mathcal J$. It is worth noting that thanks to the nearly static and LoS(-dominant) channels among the BS and all IRSs (or IRS controllers) in our considered system setup (see Fig. \ref{system_model}), the estimation of $\mu_{j}^{2},~j\in\mathcal J$ is assumed to be perfect.

On the other hand, assume that the large-scale user-associated channel gains, i.e., $\beta_{k}^{2}$ and $\alpha_{k,j}^{2},~k\in\mathcal K,j\in\mathcal J$, keep nearly constant over two consecutive transmission frames. Thus, during the direct channel estimation (i.e., Phase I in Stage II) in the previous transmission frame, the BS and IRS controller $j,~j\in\mathcal J$ can measure the average RSS from each user when all users transmit pilot symbols, which can then be used to estimate $\beta_{k}^{2}$ and $\alpha_{k,j}^{2},~k\in\mathcal K,j\in\mathcal J$ in the current transmission frame\footnote{In practice, such long-term statistical CSI can be updated in a much larger interval as compared to each frame duration, and thus is not our concern in this paper.}.

\subsection{User-IRS Association Optimization based on Statistical CSI}
Based on the estimated statistical CSI, i.e., $\beta_{k}^{2}$, $\alpha_{k,j}^{2}$, and $\mu_{j}^{2},~k\in\mathcal K,j\in\mathcal J$, in this subsection, we formulate a user-IRS association optimization problem to determine the cascaded user-IRS-BS channels that need to be estimated in Stage II. To this end, we define $0\leq \lambda_{k,j}\leq 1$ as an auxiliary variable for IRS $j$ to serve user $k$, which satisfies $\sum_{k=1}^{K}\lambda_{k,j}=1,~j\in\mathcal J$. In particular, given the maximum passive beamforming gain of $N^{2}$ by each IRS\footnote{Such a squared passive beamforming gain has been verified in the literature under various channel setups, e.g., \cite{power_scaling,correlated_Rayleigh,free_space,Rician}.} , we consider that user $k$ can reap a partial passive beamforming gain by $\lambda_{k,j}N$ (out of  the total $N$) elements\footnote{For convenience, we assume ``virtual'' fractional number of IRS elements here if $\lambda_{k,j}N$ is not an integer, which usually does not impact the association result as $N$ is practically very large and the number of selected users for each IRS $j, j\in\mathcal{J}$ (with $\lambda_{k,j}>0$) is finite.} of IRS $j$, i.e., $(\lambda_{k,j}N)^{2}$; while the remaining $N(1-\lambda_{k,j})$ reflecting elements are regarded as random scatterers for it. Note that the above is simply an approximation for the beamforming gain achievable for each user to facilitate the user-IRS association design.

Since only the statistical CSI, i.e., $\beta_{k}^{2}$, $\alpha_{k,j}^{2}$, and $\mu_{j}^{2},~k\in\mathcal K,j\in\mathcal J$, is available, to characterize the average channel condition from each user $k$ to the BS, we propose a large-scale channel metric for each user $k$ as follows,
\begin{align}
H_{k}=\beta_{k}^{2}+\sum_{j=1}^{J}\alpha_{k,j}^{2}\mu_{j}^{2}\lambda_{k,j}^{2}N^{2},~k\in\mathcal K.\label{approximation}
\end{align}
Note that $H_{k}$ accounts for the direct user $k$-BS channel gain (i.e., $\beta_{k}^{2}$), all cascaded user $k$-IRS $j$-BS channel gains (i.e., $\mu_{j}^{2}\alpha_{k,j}^{2},~j\in\mathcal J$), as well as the passive beamforming gain provided by each IRS $j,~j\in\mathcal J$ for user $k$ (i.e., $\lambda_{k,j}^{2}N^{2},~j\in\mathcal J$). It is also worth noting that in (\ref{approximation}), we ignore the random scattering by the other $N(1-\lambda_{k,j})$ elements of each IRS $j$, due to its negligible impact on the system performance in general \cite{weidong}.

Next, we optimize the auxiliary variables $\{\lambda_{k,j}\}$ to balance the users' channel metrics in (\ref{approximation}) and thereby determine the user-IRS associations based on the optimized $\{\lambda_{k,j}\}$. The corresponding optimization problem is formulated as
\begin{align}
\text{(P1):}~&\max_{\{\lambda_{k,j}\}}~\min_{k\in\mathcal K}H_{k}\nonumber\\
\text{s.t.}~&\sum_{k=1}^{K}\lambda_{k,j}=1,~j\in\mathcal J,\label{1}\\
&0\leq \lambda_{k,j}\leq 1,~j\in\mathcal J, k\in\mathcal K.\label{01}
\end{align}
Denote by $\{\lambda_{k,j}^{*}\}$ the optimal solution to (P1). Note that with a larger value of $\lambda_{k,j}^{*}$, user $k$ can reap a higher passive beamforming gain from IRS $j$. Hence, the passive reflection of IRS $j$ for data transmission in the subsequent Stage II should be designed with a higher priority for catering to the cascaded user $k$-IRS $j$-BS channel, i.e., $\mv G_{k,j}$, which thus needs to be estimated in Stage II. On the other hand, if $\lambda_{k,j}^{*}$ is small, user $k$ may barely benefit from the passive reflection of IRS $j$. Thus, there is no need to estimate its cascaded channel $\mv G_{k,j}$. Based on the above, to determine the user-IRS associations, i.e., $\delta_{k,j},~k\in\mathcal K,j\in\mathcal J$, we can sort all $\lambda_{k,j}^{*}$'s in the descending order, denoted as $\lambda_{k_{1},j_{1}}^{*}\geq\lambda_{k_{2},j_{2}}^{*}\geq\cdots\geq\lambda_{k_{KJ},j_{KJ}}^{*}$, where $\lambda_{k_{l},j_{l}}^{*}$ denotes the $l$-th largest entry, $1\leq l\leq KJ$. Then we select the $\zeta$ largest values and set their corresponding binary association variables to one, i.e., $\delta_{k_{l},j_{l}}=1,~l=1,2,\cdots,\zeta$. Thus, we only need to estimate the cascaded user $k_{l}$-IRS $j_{l}$-BS channels, $l=1,2,\cdots,\zeta$, in Stage II of the proposed transmission protocol.
\begin{remark}
 Note that we do not consider the interference among users in the user-IRS association optimization, as it only serves as an initial step to  balance the users' average channel condition with the BS when the short-term channel knowledge is not available yet. The interference suppression would be achieved by the IRS passive reflection and BS active antenna combining after the short-term cascaded/overall channel knowledge is acquired, as shown in Phases II and III in Fig. \ref{protocol}.
\end{remark}

Next, we focus on solving (P1). However, problem (P1) is a non-convex optimization problem due to its non-concave objective function, and thus it is difficult to be optimally solved. To tackle this challenge, we adopt the SCA algorithm and obtain a locally optimal solution to it. To this end, we re-express (P1) as the following epigraph form \cite{convex} by introducing an auxiliary variable $\bar{H}$, i.e,
\begin{align}
\text{(P1.1):}~&\max_{\{\lambda_{k,j}\},\bar{H}}~\bar{H}\nonumber\\
\text{s.t.}~&H_{k}\geq \bar{H},~k\in\mathcal K,\label{relax}\\
&\text{(\ref{1})~and~(\ref{01}).}\nonumber
\end{align}
Then, given a local point at each SCA iteration, we approximate the non-convex constraints in (\ref{relax}) into convex ones. By iteratively solving a sequence of approximate convex problems, we can obtain a locally optimal solution to (P1.1), and so does for (P1). 

Denote by $\{\lambda_{k,j}^{(l)}\}$ the local point at the $l$-th SCA iteration with $l\geq 1$. Notice that $H_{k}$ is a convex function with respect to $\{\lambda_{k,j}\}$; thus, it can be lower-bounded by its first-order Taylor expansion at the local point $\{\lambda_{k,j}^{(l)}\}$ in the $l$-th SCA iteration, i.e., 
\begin{align}
H_{k}&\geq \beta_{k}^{2}+\sum_{j=1}^{J}\alpha_{k,j}^{2}\mu_{j}^{2}(\lambda_{k,j}^{(l)}N)^{2}+2\sum_{j=1}^{J}(\lambda_{k,j}-\lambda_{k,j}^{(l)})(\alpha_{k,j}^{2}\mu_{j}^{2}\lambda_{k,j}^{(l)}N^{2})\triangleq H^{lb}_{k},~k\in\mathcal K.\label{lower}
\end{align}
Then, by replacing $H_{k}$ in (\ref{relax}) with $H_{k}^{lb}$ in (\ref{lower}), problem (P1.1) is approximated as the following convex optimization problem in the $l$-th SCA iteration, denoted as (P1.$l$) below, which can be optimally solved via standard convex optimization techniques (e.g., interior point method \cite{convex}). 
\begin{align}
\text{(P1.$l$):}~&\max_{\{\lambda_{k,j}\},\bar{H}}~\bar{H}\nonumber\\
\text{s.t.}~&H_{k}^{lb}\geq \bar{H},~k\in\mathcal K,\\
&\text{(\ref{1}) and (\ref{01}).}\nonumber
\end{align}
Denote the optimal solution to (P1.$l$) as $\{\lambda_{k,j}^{(l)*}\}$. Next, we update the local point in the $(l+1)$-th SCA iteration as $\lambda_{k,j}^{(l+1)}=\lambda_{k,j}^{(l)*},~k\in\mathcal K,j\in\mathcal J$. The above procedure repeats until the fractional increase in the objective value of (P1) is smaller than a pre-determined threshold $\epsilon_{1}$. We summarize the main procedure of the above SCA algorithm in Algorithm \ref{sca}. It can be shown that Algorithm \ref{sca} converges to a locally optimal solution to problem (P1.1) or (P1) \cite{sca}. Denote by $I_{s}$ the total number of SCA iterations required for Algorithm \ref{sca}. In each iteration, the convex optimization problem (P1.$l$) is solved via the standard interior-point method with the complexity of $\mathcal{O}(K^{1.5}J^{1.5})$ \cite{complexity}. As a result, the overall complexity of Algorithm \ref{sca} is $\mathcal{O}(I_{s}K^{1.5}J^{1.5})$.

\begin{algorithm}
\caption{SCA Algorithm to Solve (P1).}
\label{sca}
\begin{algorithmic}[1]
\STATE {Initialize $\lambda_{k,j}^{(1)}$ to satisfy constraint (\ref{1}) and set $l=1$.}

\REPEAT
\STATE {Given the local point $\{\lambda_{k,j}^{(l)}\}$, solve (P1.$l$) and obtain its optimal solution as $\{\lambda_{k,j}^{(l)*}\}$.}

\STATE{Update $\lambda_{k,j}^{(l+1)}=\lambda_{k,j}^{(l)*},~k\in\mathcal K,j\in\mathcal J$.}
\STATE{Update $l=l+1$.}
\UNTIL{the fractional increase in the objective value of (P1) is smaller than $\epsilon_{1}$. }
\end{algorithmic}
\end{algorithm}

\section{Short-Term Cascaded Channel Estimation and IRS Reflection Optimization}\label{ps_design}
In this section, we first present the procedure for estimating the selected cascaded user-IRS-BS channels, i.e., $\mv G_{k,j}$ with $\delta_{k,j}=1,~k\in\mathcal K,j\in\mathcal J$, in Phase II of Stage II in the proposed transmission protocol (see Fig. \ref{protocol}). Then, based on the estimates of the selected cascaded channels in Phase II (and those of all direct user-BS channels in Phase I), we jointly optimize the passive reflections of all IRSs to maximize the minimum average SINR among all users, accounting for the other cascaded channels that are not estimated.

\subsection{Short-Term Cascaded Channel Estimation}
First, based on the optimized user-IRS associations in Stage I (see Section \ref{user_IRS}-B), i.e., $\delta_{k,j},~k\in\mathcal K,j\in\mathcal J$, the BS determines the pilot sequences of all users and reflection design of all IRSs in Phase II of each transmission block in Stage II (to be specified later), which are informed to the users and IRS controllers over the downlink control links, respectively, before Stage II starts\footnote{The pilot sequences of all users in Phases I and III of each block are the same as that for the conventional multi-antenna BS without IRS.}.

Let $\mathcal{B}_{j}$ denote the set of users associated with IRS $j$, i.e., $\mathcal{B}_{j}=\{k|\delta_{k,j}=1,~k\in\mathcal K\}, j \in \cal J$. As Phase II of each block starts, the IRSs are sequentially turned on to estimate the cascaded channels for their respectively associated users. For example, when IRS $j$ is turned on, only the users in $\mathcal{B}_{j}$ will be active and transmit non-zero pilot symbols for the BS to estimate their cascaded channels via IRS $j$, i.e., $\mv G_{k,j},~k\in\mathcal{B}_{j}$. Given the reference cascaded IRS controller $U_{j}$-IRS $j$-BS channel estimated in Stage I, i.e., $\hat{\mv g}_{U_{j},j,n},~n\in\mathcal N$, the BS only needs to estimate a set of scaling factors, i.e., $\nu_{k,j,n},~k\in\mathcal {B}_{j},~n\in\mathcal N$, for which the details, including the pilot sequence of each user $k, k \in \mathcal {B}_{j}$ and the reflection design of IRS $j$ in this process, can be found in \cite{liuliang} and thus omitted here for brevity. Let $\hat{\nu}_{k,j,n}~k\in\mathcal {B}_{j},n\in\mathcal N$ denote the estimates of the above scaling factors. Then, the cascaded user $k$-IRS $j$-BS channel can be estimated as $\hat{\mv G}_{k,j}=[\hat{\mv g}_{k,j,1},\cdots,\hat{\mv g}_{k,j,N}],~k\in\mathcal{B}_{j}$, where $\hat{\mv g}_{k,j,n}=\hat{\nu}_{k,j,n}\hat{\mv g}_{U_{j},j,n},~n\in\mathcal N$. It can be shown that the associated overall training duration of Phase II is given by $\tau_{2}=\sum_{j=1}^{J}|\mathcal{B}_{j}|N/M=\zeta N/M$ \cite{liuliang}. As $\tau=\tau_{1}+\tau_{2}+\tau_{3}$, the total number of selected cascaded channels in Phase II is given by $\zeta=\lfloor(\tau-\tau_{1}-\tau_{3})M/N\rfloor$. Moreover, since $\zeta\leq KJ$, the maximum training duration of Phase II is given by $\tau_{2,\max}\triangleq KJN/M$, with which all $KJ$ cascaded user-IRS-BS channels are selected and estimated, and thus the maximum training duration in our proposed architecture is given by $\tau_{\max}=\tau_{1}+\tau_{2,\max}+\tau_{3}$.
\begin{remark}\label{bs_user}
For the conventional architecture with user-side IRSs, since the BS-IRS distance becomes considerably larger, we cannot implement Stage I as our proposed protocol for BS-side IRSs. Due to the severe double path loss of the reflected links via other remote IRSs, we only need to estimate the cascaded channel from each user to the BS via its nearby IRSs (if any) in each fading block for user-side IRSs. As a result, its minimum training duration is given by $\tau_{u,\min}\triangleq \tau_{1}+\tau_{3}+NJ+(K-J)N/M$. It then follows that $\tau_{\max}\leq \tau_{u,\min}$ if $M\geq K+1$, which can be easily achieved by BS in today's massive MIMO system with a large $M$. Furthermore, since the overall signaling overhead is proportional to the channel training duration, we can also conclude that BS-side IRSs can achieve a smaller signaling overhead as compared to user-side IRSs in general.\vspace{-9pt}
\end{remark}

\subsection{IRS Reflection Optimization based on Short-Term CSI}
In this subsection, we optimize the passive reflections of all IRSs based on the estimates of all direct channels in Phase I, i.e., $\mv h_{d,k},~k\in\mathcal K$, and selected cascaded user-IRS-BS channels in Phase II, i.e., $\hat{\mv G}_{k,j},~k\in\mathcal {B}_{j},j\in\mathcal J$. By substituting the above channel estimates into (\ref{effective}), the overall channel from user $k$ to the BS can be constructed as the superposition of two components, i.e.,
\begin{equation}\label{effective_new}
\hat{\mv h}_{k}(\tilde{\mv\theta})=(\tilde{\mv G}_{k}+\bar{\mv G}_{k})\tilde{\mv\theta}, ~k \in {\cal K},
\end{equation}
where $\tilde{\mv G}_{k}=[\delta_{k,1}\hat{\mv G}_{k,1},\cdots,\delta_{k,J}\hat{\mv G}_{k,J},\mv h_{d,k}]\in \mathbb{C}^{M\times{(NJ+1)}}$ contains the estimates of the direct and selected cascaded channels, $\bar{\mv G}_{k}=[(1-\delta_{k,1}){\mv G}_{k,1},\cdots,(1-\delta_{k,J}){\mv G}_{k,J},\mv 0]\in \mathbb{C}^{M\times{(NJ+1)}}$ contains the cascaded channels that are not estimated (thus unknown), and $\tilde{\mv\theta}=[\mv\theta_{1}^{T},\cdots,\mv\theta_{J}^{T},1]\in\mathbb{C}^{(NJ+1)\times 1}$. Note that only one of the $j$-th blocks in $\tilde{\mv G}_{k}$ and $\bar{\mv G}_{k}$, i.e., $\delta_{k,j}\hat{\mv G}_{k,j}$ and $(1-\delta_{k,j}){\mv G}_{k,j}$, would be non-zero, depending on the value of $\delta_{k,j}, j \in \cal J$. Evidently, if all cascaded channels for user $k$ are estimated, i.e., $\delta_{k,j}=1, j \in \cal J$, then we have $\bar{\mv G}_{k}={\bf 0}$. 

Due to the existence of unknown entries in $\bar{\mv G}_{k}, k \in \cal K$ in general, the IRS passive reflection cannot be optimized to ensure the instantaneous performance of all users. Nonetheless, by exploiting the available statistical CSI estimated in Stage I, i.e., $\alpha_{k,j}^{2}$ and $\mu_{j}^{2},~k\in\mathcal K,j\in\mathcal J$ (see Section \ref{user_IRS}-A), we can consider the average SINR performance of each user, i.e.,
\begin{align}
\bar{\gamma}_{k}(\mv w_{k},\tilde{\mv\theta})=\mathbb{E}\left[\frac{p|\mv w_{k}\hat{\mv h}_{k}(\tilde{\mv\theta})|^{2}}{\sum\limits_{q\neq k}p|\mv w_{k}\hat{\mv h}_{q}(\tilde{\mv\theta})|^{2}+\sigma^{2}}\right],~k\in\mathcal K,\label{average_sinr}
\end{align}
where the expectation is taken over the short-term CSI corresponding to the cascaded channels that are not estimated. However, it is generally difficult to express (\ref{average_sinr}) in a tractable form and design the  passive reflections of all IRSs accordingly. To circumvent this difficulty, we consider the following specific channel models.

Specifically, we assume that all user-IRS channels follow independent Rayleigh fading, i.e., $\mv t_{k,j}\sim\mathcal{CN}(\mv 0,\alpha_{k,j}^{2}\mv I_{N}),~k\in\mathcal K,j\in\mathcal J$, and all IRS-BS channels are dominated by a LoS link thanks to their proximity, i.e., $\mv F_{j}=\mu_{j}\mv l_{j}\tilde{\mv l}_{j}^{H},~j\in\mathcal J$, where $\mv l_{j}\in\mathbb{C}^{M\times 1}$ and $\tilde{\mv l}_{j}\in\mathbb{C}^{N\times 1}$ denote the array response at the BS and IRS $j$, respectively. Next, we approximate the average SINR in (\ref{average_sinr}) by its lower bound given below \cite{weidong}
\begin{align}
\bar{\gamma}_{k}(\mv w_{k},\tilde{\mv\theta})&=\mathbb{E}[p|\mv w_{k}\hat{\mv h}_{k}(\tilde{\mv\theta})|^{2}]\mathbb{E}\Big[\frac{1}{p\sum_{q\neq k}|\mv w_{k}\hat{\mv h}_{q}(\tilde{\mv\theta})|^{2}+\sigma^{2}}\Big]\geq\frac{p\mathbb{E}[|\mv w_{k}\hat{\mv h}_{k}(\tilde{\mv\theta})|^{2}]}{p\mathbb{E}[\sum_{q\neq k}|\mv w_{k}\hat{\mv h}_{q}(\tilde{\mv\theta})|^{2}]+\sigma^{2}}\nonumber\\
&\triangleq\tilde{\gamma}_{k}(\mv w_{k},\tilde{\mv\theta}),~k\in\mathcal K,\label{asainr}
\end{align}
where the equality is due to the fact that $\hat{\mv h}_{k}(\tilde{\mv\theta})$ and $\hat{\mv h}_{q}(\tilde{\mv\theta}),~q\neq k$ are independent, and the inequality holds due to the Jensen's inequality since the function $\frac{1}{x}$ is convex in $x$ for $x>0$; thus, we have $\mathbb{E}[1/x]\geq 1/\mathbb{E}[x]$. Then, we have the following proposition.

\begin{proposition}\label{pro_asainr}
Under the above channel models, the lower bound of the achievable SINR of user $k$ given in (\ref{asainr}) is given by
\begin{align}
&\tilde{\gamma}_{k}(\mv w_{k},\tilde{\mv\theta})=\frac{p\mv w_{k}^{H}\tilde{\mv G}_{k}\tilde{\mv\theta}\tilde{\mv\theta}^{H}\tilde{\mv G}_{k}^{H}\mv w_{k}+p\mv w_{k}^{H}\mv A_{k}\mv w_{k}}{\sum\limits_{q\neq k}\left(p\mv w_{k}^{H}\tilde{\mv G}_{q}\tilde{\mv\theta}\tilde{\mv\theta}^{H}\tilde{\mv G}_{q}^{H}\mv w_{k}+p\mv w_{k}^{H}\mv A_{q}\mv w_{k}\right)+\sigma^{2}},~k\in\mathcal K,\label{approximation_asainr}
\end{align}
where $\mv A_{k}=\sum_{j=1}^{J}(1-\delta_{k,j})N\mu_{j}^{2}\alpha_{k,j}^{2}\mv l_{j}\mv l_{j}^{H}\in\mathbb{C}^{M\times M},~k\in\mathcal K$.
\end{proposition}
\begin{IEEEproof}
See Appendix \ref{proof_asainr}.
\end{IEEEproof}

It is observed that in (\ref{approximation_asainr}), the IRS passive reflection $\tilde{\mv\theta}$ only acts on the estimated short-term CSI, i.e., $\tilde{\mv G}_{k}$, while the BS combining weights $\mv w_{k}$'s act on both long-term and short-term CSI, i.e., $\tilde{\mv G}_{k}$ and ${\mv A}_k$.

In particular, if $\tau_{2}=\tau_{2,\max}$, all cascaded user-IRS-BS channels can be estimated in Phase II, i.e., $\delta_{k,j}=1$ and ${\mv A}_k={\bf 0},~k\in\mathcal K,j\in\mathcal J$. In this case, (\ref{approximation_asainr}) becomes the instantaneous SINR of each user $k$, i.e.,
\begin{align}
\tilde{\gamma}_{k}(\mv w_{k},\tilde{\mv\theta})=\frac{p\mv w_{k}^{H}\tilde{\mv G}_{k}\tilde{\mv\theta}\tilde{\mv\theta}^{H}\tilde{\mv G}_{k}^{H}\mv w_{k}}{\sum\limits_{q\neq k}\left(p\mv w_{k}^{H}\tilde{\mv G}_{q}\tilde{\mv\theta}\tilde{\mv\theta}^{H}\tilde{\mv G}_{q}^{H}\mv w_{k}\right)+\sigma^{2}},~k\in\mathcal K.
\end{align}
In this case, the BS can directly perform a joint optimization of IRS passive reflection and BS active combining without the need of IRS-user associations.

On the other hand, if $\tau_{2}=0$, then no cascaded user-IRS-BS channel can be estimated in Phase II, i.e., $\delta_{k,j}=0,~k\in\mathcal K,j\in\mathcal J$ (and thus the reflection design is based on the long-term cascaded CSI only). As a result, (\ref{approximation_asainr}) is reduced to 
\begin{align}
&\tilde{\gamma}_{k}(\mv w_{k})=\frac{p\mv w_{k}^{H}\mv h_{d,k}\mv h_{d,k}^{H}\mv w_{k}+p\mv w_{k}^{H}\mv A_{k}\mv w_{k}}{\sum\limits_{q\neq k}\left(p\mv w_{k}^{H}\mv h_{d,q}\mv h_{d,q}^{H}\mv w_{k}+p\mv w_{k}^{H}\mv A_{q}\mv w_{k}\right)+\sigma^{2}},~k\in\mathcal K,\label{sinr_no_theta}
\end{align}
which is regardless of $\tilde{\mv\theta}$. It follows that in this case, the passive reflections of all IRSs can be randomly set without affecting the average SINRs in (\ref{sinr_no_theta}).

\begin{remark}
It is worth noting that the specific channel models considered in Proposition \ref{pro_asainr} are mainly used to simplify the analysis of user SINRs and ease the IRS passive reflection optimization. The resulting IRS reflection can still be applied if the actual channel models are different from the assumed one, e.g., Rician-fading instead of Rayleigh-fading user-IRS channels, as will be shown in Section \ref{simulation_results} via simulation results.
\end{remark}

Next, the BS optimizes the passive reflections of all IRSs to maximize the minimum lower bound of user SINR in (\ref{approximation_asainr}). The design problem is formulated as
\begin{align}
\text{(P2):}~&\max_{\{\mv w_{k}\},\tilde{\mv\theta}}\min_{k}\tilde{\gamma}_{k}(\mv w_{k},\tilde{\mv\theta})\nonumber\\
\text{s.t.}~&||\mv w_{k}||^{2}=1,~k\in\mathcal K,\label{norm1}\\
&|(\tilde{\mv\theta})_{n}|=1,~j\in\mathcal J,n=1,2,\cdots,NJ,\label{11}\\
&(\tilde{\mv\theta})_{NJ+1}=1.\label{NM+1}
\end{align}
It is worth noting that in (P2), the BS combining weights $\{\mv w_{k}\}$ are only optimized as auxiliary variables to ease the optimization of IRS passive reflection. The actual BS combining weights will be determined based on the estimates of the overall user-BS channels with designed IRSs' reflections, i.e., $\mv h_{k}(\{\mv\theta_{j}\}),~k\in\mathcal K$, as shown in (\ref{wk_final}). However, problem (P2) is a non-convex optimization problem, as its objective function is non-concave and the unit-modulus constraints in (\ref{11}) are non-convex. Moreover, the variables $\{\mv w_{k}\}$ and $\tilde{\mv\theta}$ are coupled in its objective function, which renders (P2) challenging to solve. 

To tackle the above challenges, we propose an AO algorithm to solve (P2), by alternately optimizing each of $\tilde{\mv\theta}$ and $\{\mv w_{k}\}$ with the other fixed, until convergence is achieved. Denote by $\{\mv w_{k}^{(l)}\}$ and $\tilde{\mv\theta}^{(l)}$ the optimized solutions of $\{\mv w_{k}\}$ and $\tilde{\mv\theta}$ in the $l$-th AO iteration, respectively.

{\it 1) IRS Reflection Design with Given Combining Weights:} In the $l$-th AO iteration, we first optimize the IRS passive reflection $\tilde{\mv\theta}$ given $\mv w_{k}=\mv w_{k}^{(l-1)},~k\in\mathcal K$. Then, the following problem needs to be solved,
\begin{align}
\text{(P2.1)}:~&\max\limits_{\tilde{\bm \theta}}~\min\limits_{k\in\mathcal K}~\tilde{\gamma}_{k}(\mv w_{k}^{(l-1)},\tilde{\mv\theta})\nonumber\\
\text{s.t.}~&\text{(\ref{11}) and (\ref{NM+1}).}\nonumber
\end{align}
However, (P2.1) is still challenging to be solved due to its fractional objective function. To tackle this challenge, we recast its objective function as the following subtractive form for each user $k,~k\in\mathcal K$, i.e.,
\begin{align}
&F_{k}(\mv w_{k}^{(l-1)},\tilde{\mv\theta},\tilde{t}^{(l-1)})=p\tilde{\mv\theta}^{H}\tilde{\mv G}_{k}^{H}\mv w_{k}^{(l-1)}\mv w_{k}^{(l-1)H}\tilde{\mv G}_{k}\tilde{\mv\theta}+p\mv w_{k}^{(l-1)H}\mv A_{k}\mv w_{k}^{(l-1)}\nonumber\\
&-\tilde{t}^{(l-1)}\Big(\sum_{q\neq k}\Big(p\mv w_{k}^{(l-1)H}\mv A_{q}\mv w_{k}^{(l-1)}+p\tilde{\mv\theta}^{H}\tilde{\mv G}_{q}^{H}\mv w_{k}^{(l-1)}\mv w_{k}^{(l-1)H}\tilde{\mv G}_{q}\tilde{\mv\theta}\Big)+\sigma^{2}\Big),\label{Fk}
\end{align}
where $\tilde{t}^{(l-1)}$ is an auxiliary variable and given by $\tilde{t}^{(l-1)}=\min_{k\in\mathcal K}\tilde{\gamma}_{k}(\mv w_{k}^{(l-1)},\tilde{\mv\theta}^{(l-1)})$. Based on (\ref{Fk}), instead of solving (P2.1), we solve the following optimization problem,
\begin{align}
\text{(P2.2)}:~&\max\limits_{\tilde{\bm \theta}}~\min\limits_{k\in\mathcal K}~F_{k}(\mv w_{k}^{(l-1)},\tilde{\mv\theta},\tilde{t}^{(l-1)})\nonumber\\
\text{s.t.}~&\text{(\ref{11}) and (\ref{NM+1}).}\nonumber
\end{align}
Although problem (P2.2) is still non-convex, we can utilize a low-complexity GPM \cite{convex} to solve it by following the steps below. 

Specifically, let $\Xi(\tilde{\mv\theta})$ denote the objective function of problem (P2.2), i.e., 
\begin{align}
\Xi(\tilde{\mv\theta})=\min_{k\in\mathcal K}~F_{k}(\mv w_{k}^{(l-1)},\tilde{\mv\theta},\tilde{t}^{(l-1)}).\label{Gamma}
\end{align}
To apply the GPM, our first step is to derive the gradient of (\ref{Gamma}). However, as $\Xi(\tilde{\mv\theta})$ is non-differential with respect to $\tilde{\mv\theta}$, we derive its subgradient instead, denoted as $\mv \xi(\tilde{\mv\theta})$. According to \cite{convex}, the subgradient of $\Xi(\tilde{\mv\theta})$ can be chosen as the gradient of the function that achieves the minimum value among all $F_{k}(\{\mv w_{k}^{(l-1)}\},\tilde{\mv\theta},\tilde{t}^{(l-1)}),~k\in\mathcal K$, i.e., 
\begin{align}
\mv \xi(\tilde{\mv\theta})&=\triangledown F_{k_{0}}(\mv w_{k_{0}}^{(l-1)},\tilde{\mv\theta},\tilde{t}^{(l-1)})\nonumber\\
&=2\Big(p\tilde{\mv G}_{k_{0}}^{H}\mv w_{k_{0}}^{(l-1)}\mv w_{k_{0}}^{(l-1)H}\tilde{\mv G}_{k_{0}}-\tilde{t}^{(l-1)}\Big(\sum_{q\neq k_{0}}p\tilde{\mv G}_{q}^{H}\mv w_{k_{0}}^{(l-1)}\mv w_{k_{0}}^{(l-1)H}\tilde{\mv G}_{q}\Big)\Big)\tilde{\mv\theta},\label{gradient}
\end{align}
where $k_{0}=\arg\min_{k\in\mathcal K}~F_{k}(\mv w_{k}^{(l-1)},\tilde{\mv\theta},\tilde{t}^{(l-1)})$. 

Next, we denote $\tilde{\mv\theta}^{(l,r)}$ as the updated IRS reflection vector in the $r$-th GPM iteration in solving (P2.2). Based on (\ref{gradient}), we set
\begin{align}
\tilde{\mv\theta}^{(l,r)}=\tilde{\mv\theta}^{(l,r-1)}+\rho_{r}\mv \xi(\tilde{\mv\theta}^{(l,r-1)}),\label{updatetheta1}
\end{align}
where $\rho_{r}=\rho/||\mv \xi(\tilde{\mv\theta}^{(l,r-1)})||_{2}$ is the step size in the $r$-th GPM iteration with $\rho$ being a constant step-size, and we initialize $\tilde{\mv\theta}^{(l,0)}=\tilde{\mv\theta}^{(l-1)}$. However, due to the existence of constraints (\ref{11}) and (\ref{NM+1}) in (P2.2), the updated $\tilde{\mv \theta}^{(l,r)}$ in (\ref{updatetheta1}) may not be feasible to (P2.2). As such, it should be projected onto the feasible region of (P2.2) prior to the $(r+1)$-th GPM iteration. Let ${\rm Proj}(\mv x)$ denote the (Euclidean) projection of a given point $\mv x=[x_{1},x_{2},\cdots,x_{NJ+1}]^{T}$ onto the feasible region of (P2.2), i.e., ${\rm Proj}(\mv x)\triangleq \arg\min_{\tilde{\bm\theta}}||\tilde{\mv\theta}-\mv x||$ subject to (\ref{11}) and (\ref{NM+1}). It can be shown that \cite{projection}
\begin{align}
({\rm Proj}(\mv x))_{n}=\exp\left(i\angle\left(\frac{(\mv x)_n}{(\mv x)_{NJ+1}}\right)\right), n=1,2,\cdots,NJ+1.\label{proj}
\end{align}
As a result, (\ref{updatetheta1}) is modified as 
\begin{align}
(\tilde{\mv\theta}^{(l,r)})_n\!&=\!({\rm Proj}(\tilde{\mv\theta}^{(l,r-1)}+\rho_{r}\mv \xi(\tilde{\mv\theta}^{(l,r-1)})))_{n}\!\nonumber\\
&=\!\exp\left(i\angle\left(\frac{(\tilde{\mv\theta}^{(l,r-1)}+\rho_{r}\mv \xi(\tilde{\mv\theta}^{(l,r-1)}))_{n}}{(\tilde{\mv\theta}^{(l,r-1)}+\rho_{r}\mv \xi(\tilde{\mv\theta}^{(l,r-1)}))_{NJ+1}}\right)\right).\label{updatetheta}
\end{align}
The above procedure is repeated until the fractional increase in the objective function of (P2.2) is smaller than a predefined threshold $\epsilon_{2}$. Then, we update $\tilde{\mv\theta}^{(l)}=\tilde{\mv\theta}^{(l,r-1)}$.

{\it 2) BS Combining Weights Design with Given IRS Reflection:} After optimizing $\tilde{\mv\theta}$ in the $l$-th AO iteration, we optimize the BS's combining weights $\{\mv w_{k}\}$ with $\tilde{\mv\theta}=\tilde{\mv\theta}^{(l)}$ by solving the following optimization problem,
\begin{align}
\text{(P2.3)}:~&\max\limits_{\{\mv w_{k}\}}~\min\limits_{k\in\mathcal K}~\tilde{\gamma}_{k}(\mv w_{k},\tilde{\mv\theta}^{(l)})\nonumber\\
\text{s.t.}~&\text{(\ref{norm1}).}\nonumber
\end{align}
Notice that problem (P2.3) can be decoupled into $K$ subproblems, and the $k$-th subproblem is given by
\begin{align}
\text{(P2.3.$k$)}:~&\max\limits_{\mv w_{k}}~\frac{\mv w_{k}^{H}\mv C_{k}^{(l)}\mv w_{k}}{\mv w_{k}^{H}\mv D_{k}^{(l)}\mv w_{k}}\nonumber\\
\text{s.t.}~&||\mv w_{k}||^{2}=1,\label{norm1k}
\end{align}
where 
\begin{align}
\mv C_{k}^{(l)}&=p\tilde{\mv G}_{k}\tilde{\mv\theta}^{(l)}\tilde{\mv\theta}^{(l)H}\tilde{\mv G}_{k}^{H}+p\mv A_{k},\\
\mv D_{k}^{(l)}&=\sum_{q\neq k}(p\tilde{\mv G}_{q}\tilde{\mv\theta}^{(l)}\tilde{\mv\theta}^{(l)H}\tilde{\mv G}_{q}^{H}+p\mv A_{q})+\mv I_{M}\sigma^{2}.
\end{align}
Note that if $\mv A_k={\bf 0}$, i.e., all cascaded channels are estimated in our proposed protocol, then the optimal solution to (P2.3.$k$) can be obtained as the MMSE-based combining, denoted as $\mv w_{k}^{\star}=(\mv D_{k}^{(l)})^{-1}\tilde{\mv G}_{k}\tilde{\mv\theta}^{(l)}/\lVert (\mv D_{k}^{(l)})^{-1}\tilde{\mv G}_{k}\tilde{\mv\theta}^{(l)} \rVert$. In the general case with $\mv A_k \ne {\bf 0}$, its optimal solution can be obtained by invoking the generalized Rayleigh quotient theorem \cite{pdd} as
\begin{align}
\mv w_{k}^{\star}=\frac{\mv v_{k}^{(l)}}{||\mv v_{k}^{(l)}||},~k\in\mathcal K,\label{wk}
\end{align}
where $\mv v_{k}^{(l)}$ denotes the eigenvector of $(\mv D_{k}^{(l)})^{-1}\mv C_{k}^{(l)}$ corresponding to its maximum eigenvalue. Then, we update $\mv w_{k}^{(l)}=\mv w_{k}^{\star}, k \in \cal K$. To summarize, we present the main procedure of the AO algorithm in Algorithm \ref{ao}.
\begin{proposition}\label{converge}
Algorithm \ref{ao} is able to converge to a locally optimal solution to (P2).
\end{proposition} 
\begin{IEEEproof}
See Appendix \ref{converge_proof}.
\end{IEEEproof}
\begin{algorithm}
\caption{Proposed AO Algorithm to Solve (P2).}
\label{ao}
\begin{algorithmic}[1]
\STATE {Initialize $\{\mv w_{k}^{(0)}\}$, $\tilde{\mv\theta}^{(0)}$ and $l=1$.}
\REPEAT
\STATE {Given $\{\mv w_{k}^{(l-1)}\}$ and $\tilde{\mv\theta}^{(l-1)}$, update $\tilde{t}^{(l-1)}=\min_{k\in\mathcal K}\tilde{\gamma}_{k}(\mv w_{k}^{(l-1)},\tilde{\mv\theta}^{(l-1)})$. 
\STATE Let $r=1$ and initialize $\tilde{\mv\theta}^{(l,r-1)}=\tilde{\mv\theta}^{(l-1)}$. }
\REPEAT
\STATE {Update $\tilde{\mv\theta}^{(l,r)}$ based on (\ref{updatetheta}).}
\STATE {Update $r=r+1$.}
\UNTIL{the fractional increase in the objective value of (P2.2) is smaller than $\epsilon_{2}$.}
\STATE{Update $\tilde{\mv\theta}^{(l)}=\tilde{\mv\theta}^{(l,r-1)}$.}
\STATE {Given $\tilde{\mv\theta}^{(l)}$, update each $\mv w_{k}^{(l)}, k \in \cal K$ as $\mv w_{k}^{\star}$ in (\ref{wk}).}
\STATE{Update $l=l+1$.}
\UNTIL{the fractional increase in the objective value of (P2) is smaller than $\epsilon_{3}$. }
\end{algorithmic}
\end{algorithm}

The complexity of Algorithm \ref{ao} is mainly from two steps, i.e., solving (P2.2) via the GPM and updating $\{\mv w_{k}\}$ based on (\ref{wk}). The complexity of the former is given by $\mathcal{O}(I_{in}(KNJM+KM^{2}+N^{2}J^{2}))$, where $I_{in}$ denotes the total number of GPM iterations. While the complexity of the latter is mainly due to the eigenvalue decomposition and matrix inverse, both of which incur a complexity of $\mathcal{O}(M^{3})$. As a result, the overall complexity of Algorithm \ref{ao} is given by $\mathcal{O}(I_{out}(I_{in}(KNJM+KM^{2}+N^{2}J^{2})+M^{3}))$, where $I_{out}$ denotes the total number of AO iterations.

\section{Numerical Results}\label{simulation_results}
\begin{figure}
\centering
\includegraphics[width=6cm]{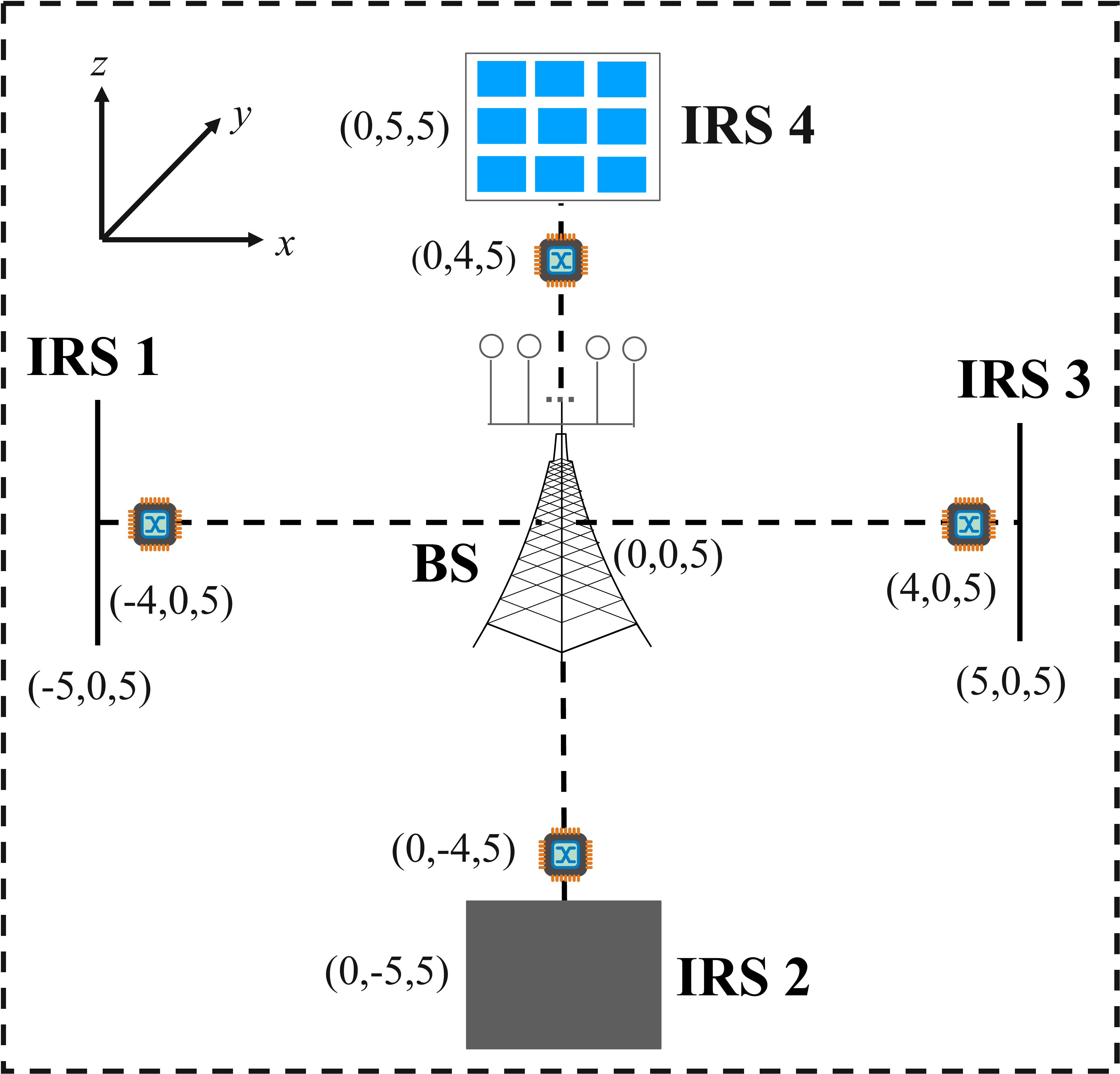}
\caption{Simulation setup.}\label{stepup}
\end{figure}
In this section, we provide numerical results to validate the performance of our proposed transmission protocol and its pertaining designs. Unless otherwise specified, the simulation parameters are set as follows. As shown in Fig. \ref{stepup}, we consider a wireless network with one BS with $M=10$ antennas, and $J=4$ co-site IRSs, each of which has $N=200$ reflecting elements. We assume a uniform planar array (UPA) at each IRS with half-wavelength spacing between any two adjacent IRS elements, and a uniform linear array (ULA) at the BS with half-wavelength antenna spacing. The coordinates of the BS, all IRSs, and all IRS controllers are shown in Fig. \ref{stepup}. We assume that there are $K=6$ users, whose locations are randomly generated within the region $[-100~\text{m},100~\text{m}]\times[-100~\text{m},100~\text{m}]$. The models and parameters of all involved channels are given in Table \ref{channel_parameter}. The transmit power of each user is set to $p=30~\text{dBm}$ in both the data and pilot transmission in Stage II, while the transmit power of each IRS controller is set to $43~\text{dBm}$ in Stage I, as shown in Fig. \ref{protocol}. The path-loss at a reference distance of $\text{1 m}$ is set to $\text{--30 dB}$ and the noise power is $\sigma^{2}=-80~\text{dBm}$.  The step size of the gradient projection method in Algorithm \ref{ao} is set as $\rho=0.01$. The stopping thresholds in Algorithms \ref{sca} and \ref{ao} are set to $\epsilon_{1}=10^{-5}$, $\epsilon_{2}=10^{-3}$, and $\epsilon_{3}=10^{-3}$, respectively. In Algorithm \ref{sca}, the auxiliary variables are initialized as $\lambda_{k,j}^{(1)}=\mu_{j}^{2}\alpha_{k,j}^{2}/(\sum_{k=1}^{K}\mu_{j}^{2}\alpha_{k,j}^{2}),~k\in\mathcal K,j\in\mathcal J$, which is proportional to the average channel gain of the corresponding cascaded channel $\mv G_{k,j}$. While in Algorithm \ref{ao}, the initial phase shifts of all IRSs are randomly generated within the interval $[0,2\pi)$, and the BS's combining weights are initialized based on (\ref{wk}). Furthermore, we set $T_{u}=5000$, $\Gamma=8~\text{dB}$, and $\tau_{1}=\tau_{3}=2K$. All results are averaged over 100 independent realizations of random channels and user locations.
\begin{table}
\centering
\caption{Channel Model and Parameter for Simulation}\label{channel_parameter}
\begin{tabular}{|p{3.1cm}|p{1cm}|p{1.7cm}|p{1.5cm}|}
\hline
&Channel model&Path-loss exponent&Rician $K$ factor (dB)\\
\hline
User-BS channel &Rician & 3.5&3\\
\hline
IRS-BS channel &Rician & 2.1&30\\
\hline
User-IRS channel (outside IRS reflection half-space)&Rayleigh& 4.8&$-\infty$\\
\hline
User-IRS channel (inside IRS reflection half-space)&Rician & 3&3\\
\hline
IRS controller-IRS channel&Rician&2.1&30\\
\hline
\end{tabular}
\end{table}

To more flexibly balance the channel training time for cascaded channel estimation in Phase II of Stage II and the users' SINRs, we further apply the IRS element grouping strategy \cite{ofdm} in the simulation, where adjacent IRS elements are grouped into a subsurface and employ a common reflection coefficient. Accordingly, the estimated element-level cascaded channels in our proposed protocol become their subsurface-level counterparts, which thus incur shorter channel training time due to their smaller numbers, but at the cost of reduced passive reflection gain for each IRS \cite{ofdm}. Let $N_{1}$ denote the number of groups (or subsurfaces) per IRS in the considered grouping strategy. As a result, the group size (or number of reflecting elements per subsurface) is $N/N_{1}$, which is assumed to be an integer. It follows that the total number of selected cascaded channels and the maximum training duration of Phase II, as given at the end of Section \ref{ps_design}-A, can be increased and reduced to $\zeta=\lfloor(\tau-\tau_{1}-\tau_{3})M/N_{1}\rfloor$ and $\tau_{2,\max}=KJN_{1}/M$, respectively. In the simulation, we set $N_{1}=50$, with which the maximum training duration of Stage II is given by ${\tau}_{\max}=\tau_{1}+\tau_{2,\max}+\tau_{3}=2K+KJN_{1}/M+2K=144$.

\subsection{User-IRS Association Design}
First, we show the efficacy of our proposed user-IRS association design in Section \ref{user_IRS}-B by comparing it with the following two benchmarks.
\begin{itemize}
\item {\bf Greedy selection:} In this benchmark, given the large-scale channel gain of all $KJ$ cascaded channels, i.e., $\mu_{j}^{2}\alpha_{k,j}^{2},~k\in\mathcal K,j\in\mathcal J$, we select $\zeta$ cascaded channels with the largest average cascaded channel gains to estimate.
\item {\bf Random selection:} In this benchmark, we randomly choose $\zeta$ cascaded channels to estimate.
\end{itemize}
Fig. \ref{user2} shows the max-min achievable rate in (\ref{rate}) among all users versus the training duration $\tau$ by different user-IRS association designs. It is observed from Fig. \ref{user2} that if only a subset of all cascaded channels are estimated, i.e., $\tau<{\tau}_{\max}$, our proposed user-IRS association design achieves better performance than the two benchmarks, as it balances the channel conditions of all users, instead of the two benchmarks. Whereas if all cascaded channels are estimated, i.e., $\tau = {\tau}_{\max}$, all considered association designs are observed to achieve the same performance, as expected.

\subsection{Effect of Training Duration in Stage II}
\begin{figure}
\centering
\begin{minipage}[t]{0.48\textwidth}
\centering
\includegraphics[width=7cm]{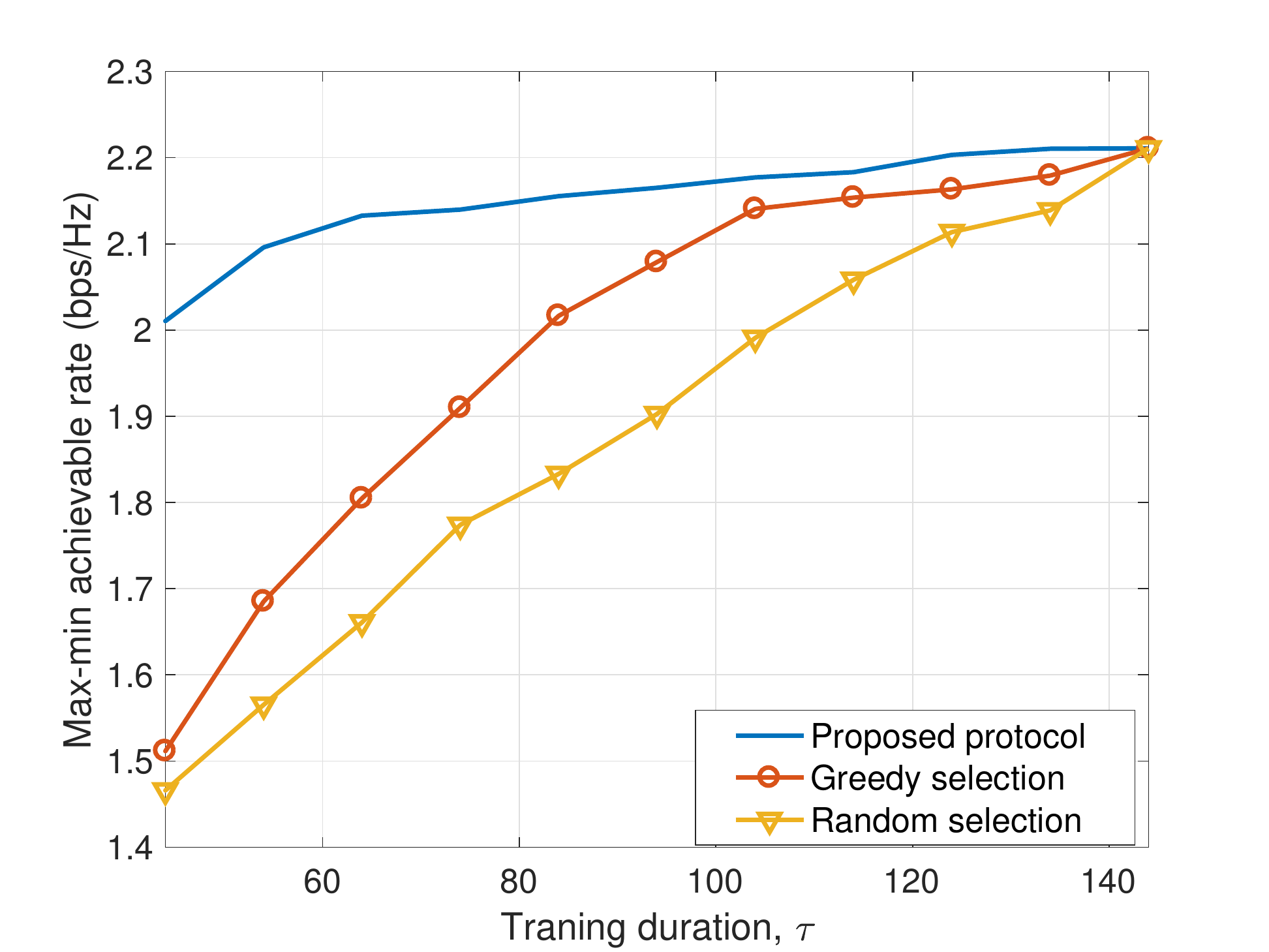}
\caption{Max-min achievable rate versus training duration, $\tau$ under different user-IRS association designs.}\label{user2}
\end{minipage}
\begin{minipage}[t]{0.48\textwidth}
\centering
\includegraphics[width=7cm]{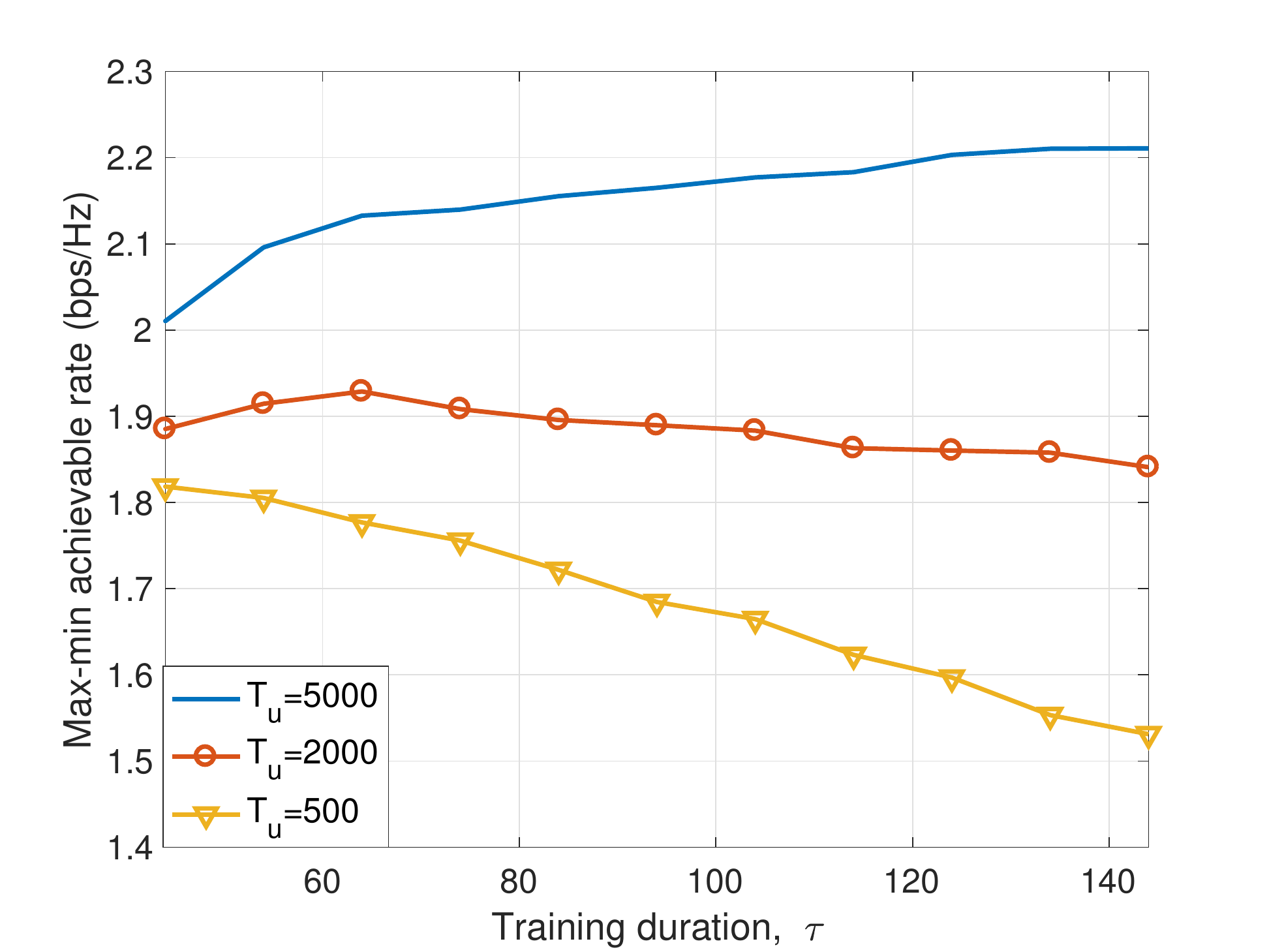}
\caption{Max-min achievable rate versus training duration, $\tau$ under different length of fading block, $T_{u}$.}\label{tradeoff}
\end{minipage}
\end{figure}
Next, we investigate the impact of the training duration $\tau$ in Stage II on the performance of the proposed protocol. In Fig. \ref{tradeoff}, we plot the max-min achievable rate in (\ref{rate}) among all users versus $\tau$ under different length of each fading block, i.e., $T_{u}$. It is observed that when $T_{u}$ is small, i.e., $T_{u}=500$, the max-min achievable rate decreases with increasing $\tau$, due to the more dominant effect of the data transmission time on the user achievable rates in (\ref{rate}) than the user SINRs in (\ref{sinr_k}). However, when $T_{u}$ is sufficiently large, i.e., $T_{u}=5000$, it is observed that the max-min achievable rate monotonically increases with $\tau$. The reason is that in this case, the considered training duration $\tau$ is much smaller than $T_{u}$. As a result, the users' achievable rates in (\ref{rate}) are more dominated by their SINRs in (\ref{sinr_k}), which can be improved as more cascaded channels are estimated. Finally, when $T_{u}=2000$, it is observed that the max-min achievable rate first increases with $\tau$ when $\tau\leq 62$ and then decreases with it. This indicates that in the case of a moderate $T_u$, the data transmission time and user SINRs in (\ref{sinr_k}) should be optimally balanced to maximize the user achievable rates in (\ref{rate}). Our proposed protocol provides flexibility to balance between these two factors in maximizing the users' achievable rates.

\subsection{Performance Comparison}
\begin{figure}
\centering
\begin{minipage}[t]{0.48\textwidth}
\centering
\includegraphics[width=7cm]{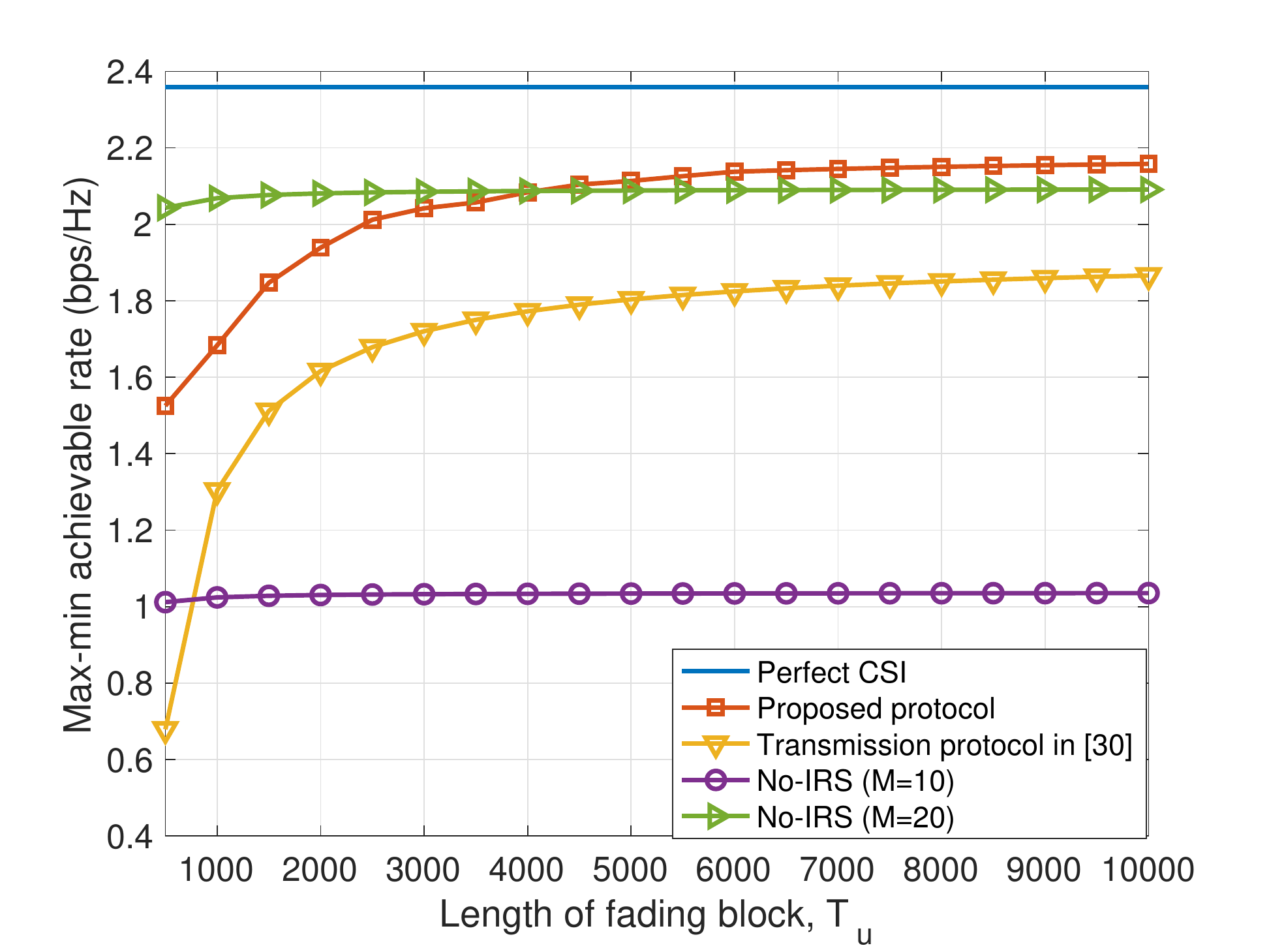}
\caption{Max-min achievable rate versus fading block length, $T_{u}$.}\label{T}
\end{minipage}
\begin{minipage}[t]{0.48\textwidth}
\centering
\includegraphics[width=7cm]{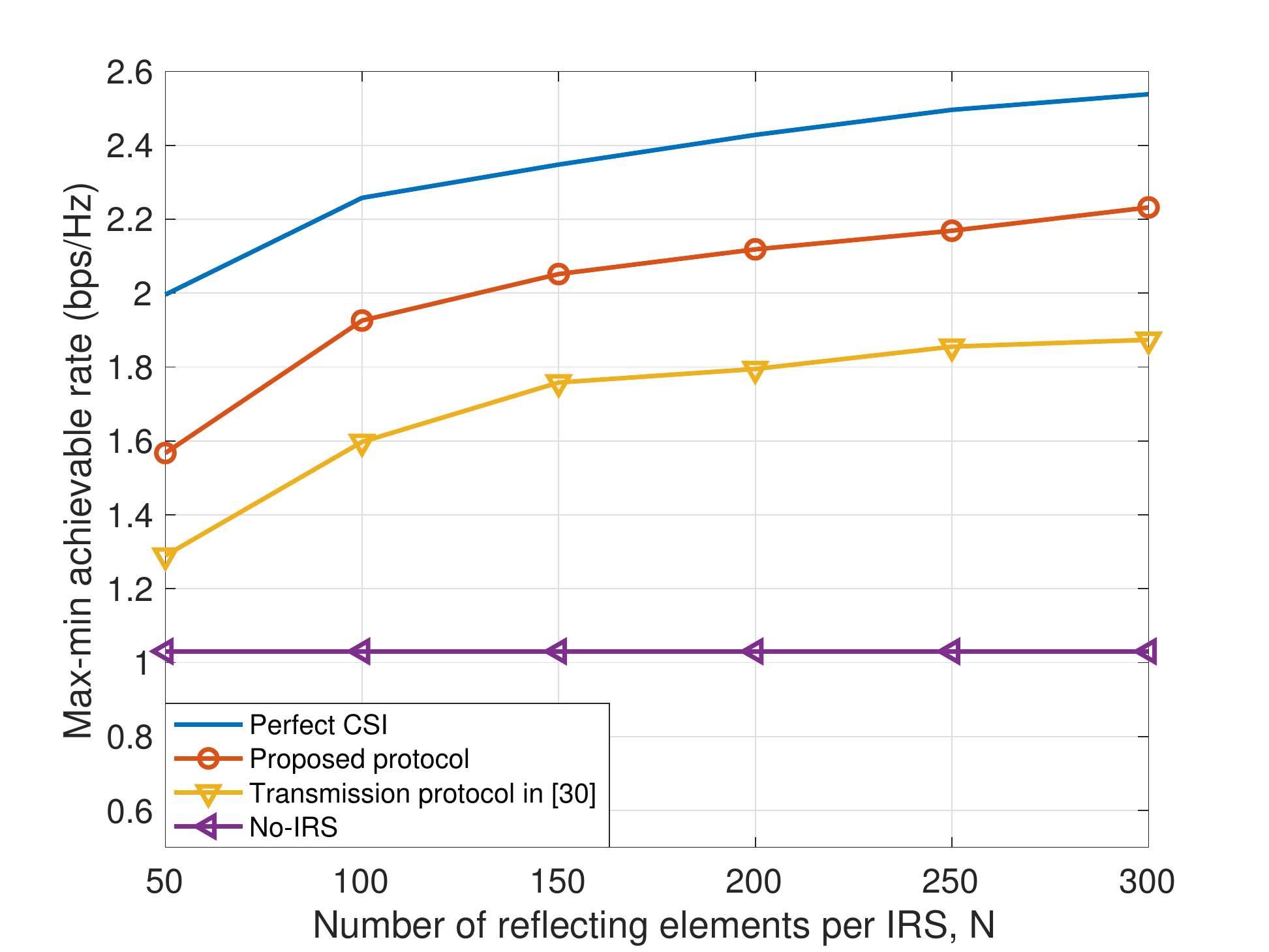}
\caption{Max-min achievable rate versus number of reflecting elements per IRS, $N$.}\label{N}
\end{minipage}
\end{figure}

Next, we show the efficacy of the proposed transmission protocol in Fig. \ref{protocol} by comparing it with the following three benchmark protocols.
\begin{itemize}
\item {\bf Transmission protocol in \cite{liuliang}:} In this transmission protocol, we apply the channel estimation method designed for {\it user-side IRSs} in \cite{liuliang}. Specifically, when each IRS $j,~j\in\mathcal J$ is turned on, one user (instead of an IRS controller in the proposed protocol) is randomly selected as the reference user $U_{j}$ to assist in the cascaded channel estimation for the remaining $K-1$ users. Since the user-IRS channels vary over different blocks, the cascaded channels of the reference users need to be updated in each fading block. As a result, there is no Stage I required and its total training duration for each fading block is equal to $\tau_{1}+\tau_{3}+N_{1}J+J(K-1)N_{1}/M=324$ \cite{liuliang}, which is even much longer than the maximum training duration of Stage II in our proposed protocol, $\tau_{\max}=144$.
\item {\bf Perfect CSI:} In this case, we assume that the BS is aware of the perfect CSI on all involved links and is able to optimize the IRS reflection and its combining weights directly without the need of channel training. Thus, this benchmark serves as an upper bound of the performance of any practical transmission protocol.
\item {\bf No-IRS:} In this case, we consider the conventional multi-antenna BS without co-site IRS, and it determines the MMSE combining based on its estimates of the direct channels with the users. Accordingly, there is no Stage I required and the required training duration for each fading block is $\tau_{1}$ only.
\end{itemize} 
Fig. \ref{T} shows the max-min achievable rate in (\ref{rate}) among all users versus $T_{u}$ by all considered schemes, where we set the training duration as $\tau=\min({\tau}_{\max},T_{u}/50)$. It is observed that the max-min achievable rate by our proposed protocol first increases with $T_{u}$ and then keeps almost constant when $T_{u}>7200$. This is because when $T_{u}<7200$, we have $T_{u}/50<144={\tau}_{\max}$. As such, an increasing number of cascaded channels can be estimated with increasing $T_u$, which helps refine the user SINRs in (\ref{sinr_k}). Due to the fixed time ratio of data transmission, i.e., $(T_{u}-\tau)/T_{u}=49/50$, the user achievable rates in (\ref{rate}) can be increased, until $T_{u}=7200$ is reached. It is also observed that there exists a gap between our proposed protocol and the benchmark scheme with perfect CSI even when $T_u$ is very large. This is owing to the channel estimation error in Phase II of Stage II in our proposed protocol and its lower performance gain of IRS passive reflection due to the IRS element grouping. In addition, it is observed that under all $T_{u}$ considered, our proposed protocol outperforms the transmission protocol in \cite{liuliang}, especially when $T_{u}$ is small. This is attributed to the following two reasons. First, as previously mentioned, the protocol in \cite{liuliang} consumes much longer training time than ours, thus severely reducing the data transmission time. Second, its random selection of reference users may result in accumulated channel estimation error for the other users if the channels of the reference user is not accurately estimated. This may happen, for example, if the reference user is selected as a user outside an IRS's reflection half-space, as the user-IRS channel may be severely attenuated by the environment scattering. While in our proposed protocol, this issue can be avoided by letting IRS controllers serve as equivalent reference users thanks to the closely deployed IRSs near the BS. Finally, it is observed that our proposed protocol significantly outperforms the no-IRS benchmark scheme employing the same number of antennas at the BS ($M=10$), despite its longer channel training time for real-time cascaded channel estimation. In fact, even if the number of antennas at the BS is doubled ($M=20$) in the latter scheme, the former scheme can still achieve a better performance than it when $T_u \ge 4000$ (or $\tau=T_u/50 \ge 80$), which implies its effectiveness in reducing the number of active antennas required for the conventional BS without co-site IRS. This manifests the great potential of co-site-IRS empowered BS as a low-cost and high-performance BS design in future wireless networks.

\subsection{Effect of Number of IRS Reflecting Elements}
Furthermore, we study the impact of the number of reflecting elements per IRS $N$ on the performance of different schemes. In Fig. \ref{N}, we plot the max-min achievable rate in (\ref{rate}) among all users versus $N$. The training duration $\tau$ is set to estimate all direct and overall user-BS channels in Phases I and III of Stage II, respectively, but only half of all cascaded channels in Phase II of Stage II, i.e., $\tau=\tau_{1}+\tau_{3}+\lceil\tau_{2,\max}/2\rceil$. It is observed from Fig. \ref{N} that our proposed protocol still significantly outperforms the transmission protocol in \cite{liuliang} and the no-IRS benchmark. Moreover, its max-min achievable rate monotonically increases with $N$, although increasing $N$ seems to incur a longer channel training duration and may greatly reduce the data transmission time. This is because by utilizing the IRS element grouping strategy, the data transmission time only depends on the fixed number of groups (subsurfaces) $N_{1}=50$ and thus remains unchanged as $N$ increases, while the user SINRs in (\ref{sinr_k}) can be improved as the group size, i.e., $N/N_{1}$, increases thanks to the larger aperture gain provided by each subsurface. Hence, the user achievable rates in (\ref{rate}) can be improved with $N$. This demonstrates that the IRS element grouping strategy can be an effective means to ensure the performance scalability of the co-site-IRS empowered BS with increasing $N$.
\begin{table}
\centering
\caption{Channel Model and Parameter for User-Side IRSs}\label{channel_parameter2}
\begin{tabular}{|p{3.1cm}|p{1cm}|p{1.7cm}|p{1.5cm}|}
\hline
&Channel model&Path-loss exponent&Rician $K$ factor (dB)\\
\hline
User-BS channel &Rician & 3.5&3\\
\hline
IRS-BS channel &Rician & 3&3\\
\hline
User-IRS channel (remote)&Rayleigh& 4.8&$-\infty$\\
\hline
User-IRS channel (nearby)&Rician & 2.1&30\\
\hline
\end{tabular}
\end{table}

\subsection{Performance Comparison Between BS-Side IRSs and User-Side IRSs}
\begin{figure}
\centering
\begin{minipage}[t]{0.48\textwidth}
\centering
\includegraphics[width=7cm]{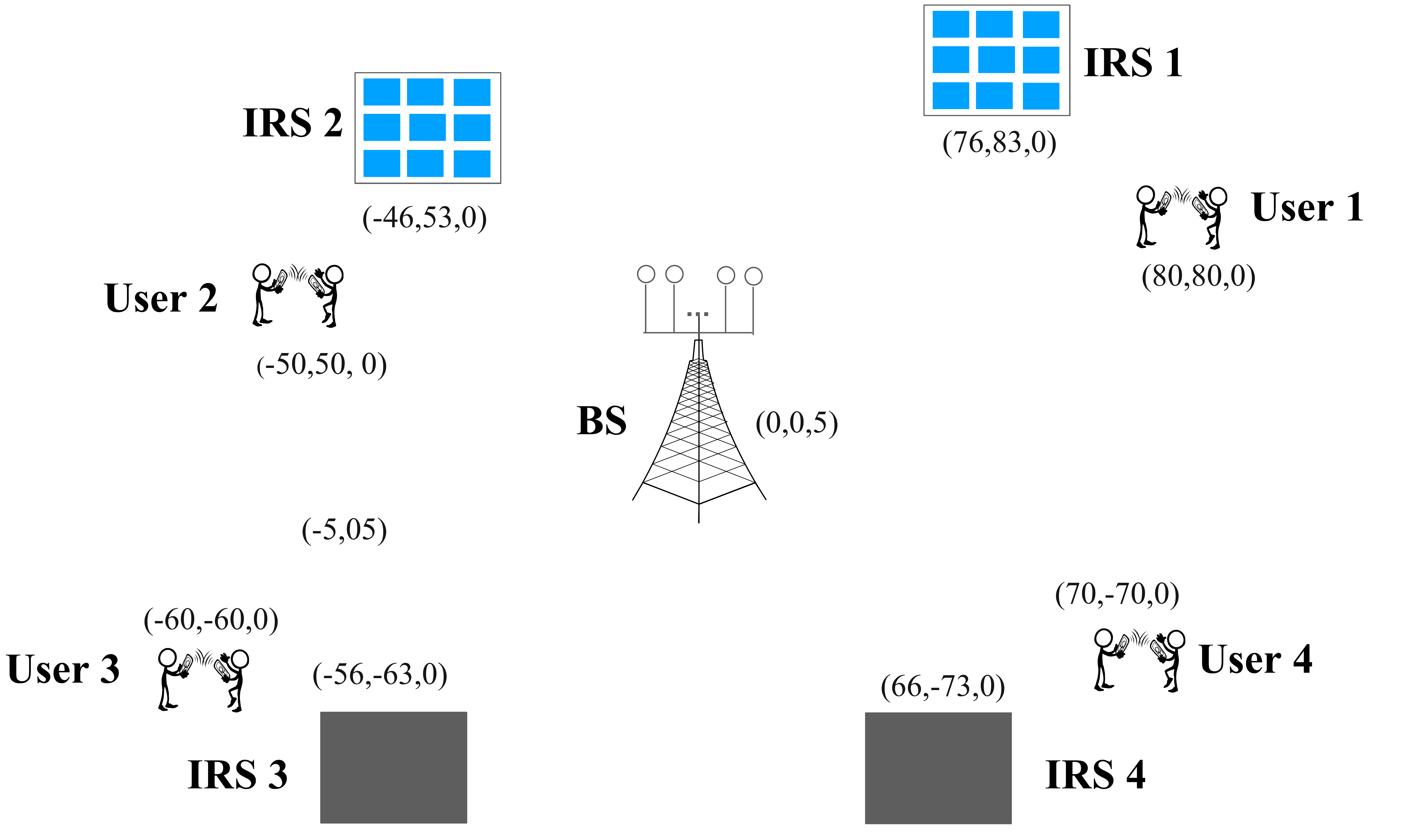}
\caption{Simulation setup for user-side IRSs.}\label{user_side_setup}
\end{minipage}
\begin{minipage}[t]{0.48\textwidth}
\centering
\includegraphics[width=7cm]{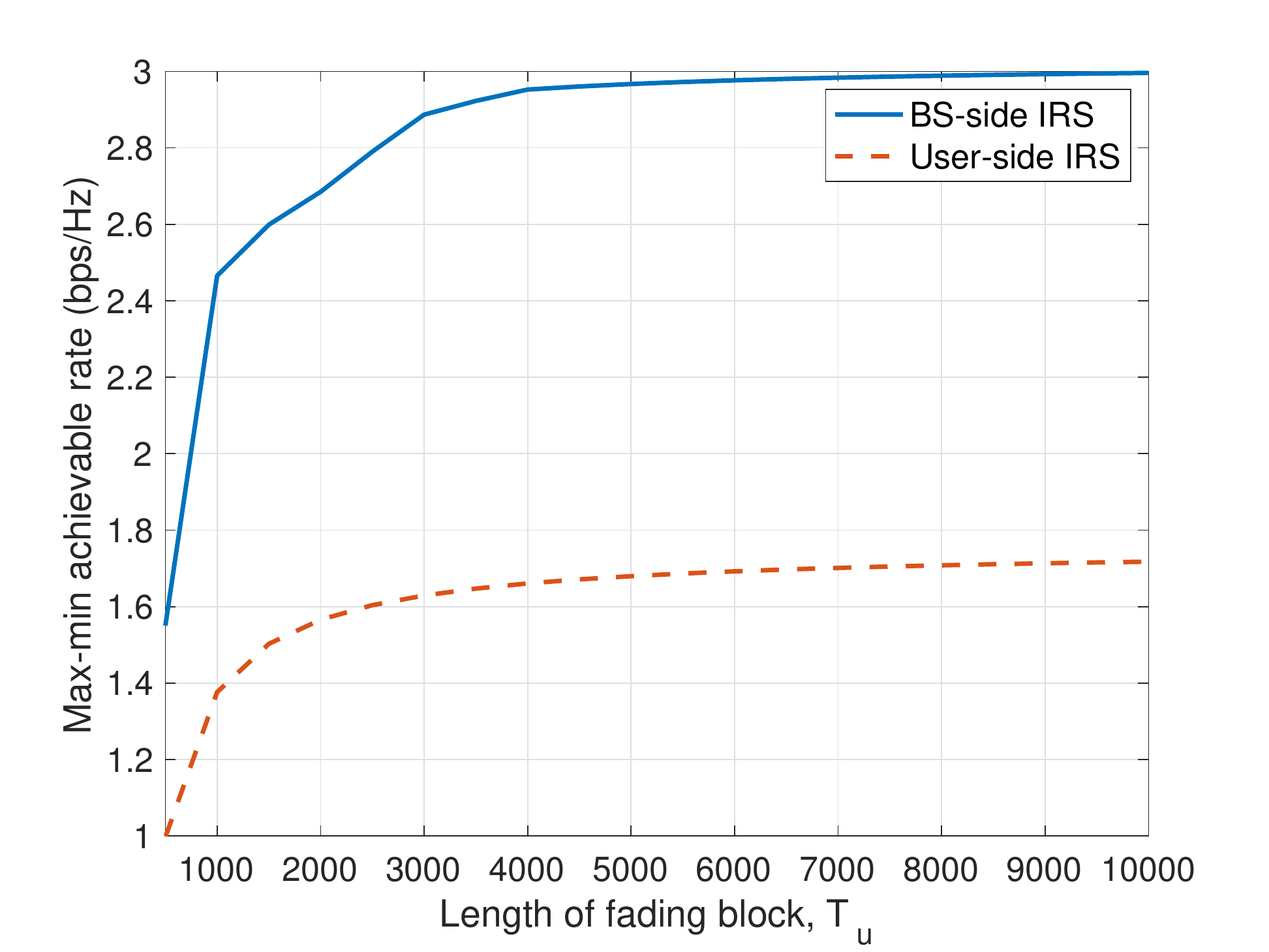}
\caption{Performance comparison between BS-side IRSs and user-side IRSs.}\label{compare_side}
\end{minipage}
\end{figure}

Finally, we compare the performance between the architectures with BS-side IRSs and user-side IRSs with $K=4$ users and $J=4$ IRSs. For the former architecture, we apply the same IRS locations and channel models as  in Fig. \ref{stepup} and in Table \ref{channel_parameter}, respectively. While for the latter architecture, we consider the simulation setup as shown in Fig. \ref{user_side_setup}. Specifically, we consider that there exists one IRS near each user and their distance is set as $5~\text{m}$ (which is the same as the BS-IRS distance in Fig. \ref{stepup}). The models and parameters of all involved channels for user-side IRSs are given in Table \ref{channel_parameter2}. It follows from Remark \ref{bs_user} that its total training duration is given by $\tau_{u,\min}=\tau_{1}+\tau_{3}+NJ=216$, which is still much longer than the maximum training duration of our proposed architecture with BS-side IRSs, i.e., $\tau_{\max}=104$.

Fig. \ref{compare_side} shows the max-min achievable rates among all users versus $T_{u}$ by the two architectures, respectively. The training duration of our proposed architecture is set as $\tau=\min(\tau_{\max},T_{u}/50)$. It is observed that our proposed architecture with BS-side IRSs achieves much better performance than that with user-side IRSs under all considered  $T_{u}$. On one hand, this is because each user can only be served by the reflecting elements of one IRS with the user-side IRSs, while it can be served by more reflecting elements with the BS-side IRSs. On the other hand, this is due to the much smaller channel training duration of the proposed architecture, as previously mentioned.

\section{Conclusions}
In this paper, we consider a new co-site-IRS empowered BS architecture and propose a practical two-stage transmission protocol for its associated IRS channel estimation and passive reflection optimization. In the first stage, we propose a cooperative channel training scheme for  the BS and IRS controllers to estimate useful long-term CSI, based on which a user-IRS association optimization problem is formulated and solved to determine the set of cascaded channels which need to be estimated in real time in the second stage. In the second stage, we present the method for the real-time cascaded channel estimation and jointly optimize the passive reflections of all IRSs to maximize the minimum average SINR among all users. 

Numerical results demonstrate that the proposed transmission protocol achieves better performance than the benchmark protocol designed for user-side IRSs. It is also shown that the considered co-site-IRS empowered BS can effectively reduce the number of active antennas required for the conventional BS without co-site IRS. Finally, the IRS element grouping strategy is shown as a scalable solution for the co-site-IRS empowered BS as the number/size of IRSs increases. There are several promising directions worthy of further investigation for the co-site-IRS empowered BS in future work, such as its channel estimation and performance optimization in the more general wide-band system, practical IRS phase-shift model \cite{practical_ps}, etc. Our proposed IRS-empowered BS architecture can also be integrated into the future cell-free system by deploying multiple IRSs near each cooperating BS to further enhance its performance. In addition, it is interesting to investigate the hybrid IRS deployment with both BS- and user-side IRSs for combining their complementary advantages. Moreover, how to take the inter-IRS signal reflections into consideration is also worthy of an in-depth investigation.

\vspace{-15pt}\appendix
\subsection{Proof of Proposition \ref{pro_asainr}}\label{proof_asainr}
First, by expanding $\mathbb{E}[|\mv w_{k}\hat{\mv h}_{l}(\tilde{\mv\theta})|^{2}],~k,l\in\mathcal K$, we have
\begin{align}
\mathbb{E}[|\mv w_{k}\hat{\mv h}_{l}(\tilde{\mv\theta})|^{2}]&=\mv w_{k}^{H}\tilde{\mv G}_{l}\tilde{\mv\theta}\tilde{\mv\theta}^{H}\tilde{\mv G}_{l}^{H}\mv w_{k}+\mv w_{k}^{H}\mathbb{E}[\tilde{\mv G}_{l}\tilde{\mv\theta}\tilde{\mv\theta}^{H}\bar{\mv G}_{l}^{H}]\mv w_{k}+\mv w_{k}^{H}\mathbb{E}[\bar{\mv G}_{l}\tilde{\mv\theta}\tilde{\mv\theta}^{H}\bar{\mv G}_{l}^{H}]\mv w_{k}.\label{asainr_appendix}
\end{align}

Then, under the considered specific channel models in Proposition \ref{pro_asainr}, we have 
\begin{align}
\mathbb{E}[\tilde{\mv G}_{l}\tilde{\mv\theta}\tilde{\mv\theta}^{H}\bar{\mv G}_{l}^{H}]&=\sum_{j=1}^{J}\sum_{n=1}^{N}\delta_{k,j}(1-\delta_{k,j})\mu_{j}^{2}\mv l_{j}\mv l_{j}^{H}\mathbb{E}[t_{k,j,n}t_{k,j,n}^{*}]=\sum_{j=1}^{J}\delta_{k,j}(1-\delta_{k,j})N\mu_{j}^{2}\alpha_{k,j}^{2}\mv l_{j}\mv l_{j}^{H},\nonumber\\
\mathbb{E}[\bar{\mv G}_{l}\tilde{\mv\theta}\tilde{\mv\theta}^{H}\bar{\mv G}_{l}^{H}]&=\sum_{j=1}^{J}\sum_{n=1}^{N}(1-\delta_{k,j})^{2}\mu_{j}^{2}\mv l_{j}\mv l_{j}^{H}\mathbb{E}[t_{k,j,n}t_{k,j,n}^{*}]=\sum_{j=1}^{J}(1-\delta_{k,j})^{2}N\mu_{j}^{2}\alpha_{k,j}^{2}\mv l_{j}\mv l_{j}^{H}.\nonumber
\end{align}
By utilizing the facts that $\delta_{k,j}(1-\delta_{k,j})=0$ and $(1-\delta_{k,j})^{2}=1-\delta_{k,j}$ in the above, we have
\begin{align}
\mathbb{E}[\tilde{\mv G}_{l}\tilde{\mv\theta}\tilde{\mv\theta}^{H}\bar{\mv G}_{l}^{H}]&=0,\label{0}\\
\mathbb{E}[\bar{\mv G}_{l}\tilde{\mv\theta}\tilde{\mv\theta}^{H}\bar{\mv G}_{l}^{H}]&=\sum_{j=1}^{J}(1-\delta_{k,j})N\mu_{j}^{2}\alpha_{k,j}^{2}\mv l_{j}\mv l_{j}^{H}.\label{no0}
\end{align}
By substituting (\ref{0}) and (\ref{no0}) into (\ref{asainr_appendix}), we obtain
\begin{align}
\mathbb{E}[|\mv w_{k}\hat{\mv h}_{l}(\tilde{\mv\theta})|^{2}]&=\mv w_{k}^{H}\tilde{\mv G}_{l}\tilde{\mv\theta}\tilde{\mv\theta}^{H}\tilde{\mv G}_{l}^{H}\mv w_{k}+\mv w_{k}^{H}\sum_{j=1}^{J}(1-\delta_{k,j})N\mu_{j}^{2}\alpha_{k,j}^{2}\mv l_{j}\mv l_{j}^{H}\mv w_{k}.\label{asainr_appendix2}
\end{align}
Then, substituting (\ref{asainr_appendix2}) into (\ref{asainr}) yields (\ref{approximation_asainr}) in Proposition \ref{pro_asainr}.The proof is thus completed.

\vspace{-10pt}\subsection{Proof of Proposition \ref{converge}}\label{converge_proof}
First, we show that with any given $\{\mv w_{k}^{(l-1)}\}$, the objective value of (P2.1) is non-decreasing by updating $\tilde{\mv\theta}^{(l-1)}$ as $\tilde{\mv\theta}^{(l)}$. Specifically, by substituting $\tilde{t}^{(l-1)}=\min_{k\in\mathcal K}\tilde{\gamma}_{k}(\mv w_{k}^{(l-1)},\tilde{\mv\theta}^{(l-1)})$ and $\tilde{\mv\theta}=\tilde{\mv\theta}^{(l-1)}$ into (\ref{Fk}), we have
\begin{align}
\min_{k\in\mathcal K}F_{k}(\mv w_{k}^{(l-1)},\tilde{\mv\theta}^{(l-1)},\tilde{t}^{(l-1)})=0.
\end{align}
Since $\tilde{\mv\theta}^{(l)}$ is an optimal solution to (P2.2), it must hold that
\begin{align}
\min_{k\in\mathcal K}F_{k}(\mv w_{k}^{(l-1)},\tilde{\mv\theta}^{(l)},\tilde{t}^{(l-1)})&\geq \min_{k\in\mathcal K}F_{k}(\mv w_{k}^{(l-1)},\tilde{\mv\theta}^{(l-1)},\tilde{t}^{(l-1)})=0,
\end{align}
which is equivalent to 
\begin{align}
\min_{k\in\mathcal K}\tilde{\gamma}_{k}(\mv w_{k}^{(l-1)},\tilde{\mv\theta}^{(l)})\geq \tilde{t}^{(l-1)}=\min_{k\in\mathcal K}\tilde{\gamma}_{k}(\mv w_{k}^{(l-1)},\tilde{\mv\theta}^{(l-1)}).\label{theta_convergence}
\end{align}

Next, it follows from (\ref{wk}) that 
\begin{align}
\min_{k\in\mathcal K}\tilde{\gamma}_{k}(\mv w_{k}^{(l)},\tilde{\mv\theta}^{(l)})\geq\min_{k\in\mathcal K}\tilde{\gamma}_{k}(\mv w_{k}^{(l-1)},\tilde{\mv\theta}^{(l)}),\label{w_convergence}
\end{align}
which indicates that the objective value of (P2.3) is non-decreasing.

Finally, by combining (\ref{theta_convergence}) and (\ref{w_convergence}), we obtain
\begin{align}
\min_{k\in\mathcal K}\tilde{\gamma}_{k}(\mv w_{k}^{(l)},\tilde{\mv\theta}^{(l)})&\geq\min_{k\in\mathcal K}\tilde{\gamma}_{k}(\mv w_{k}^{(l-1)},\tilde{\mv\theta}^{(l)})\geq \min_{k\in\mathcal K}\tilde{\gamma}_{k}(\mv w_{k}^{(l-1)},\tilde{\mv\theta}^{(l-1)}),
\end{align}
which indicates that the objective value of (P2) is non-decreasing for any two consecutive AO iterations. As the optimal value of (P2) is bounded above, Algorithm \ref{ao} is able to converge to a locally optimal solution to (P2). The proof is thus completed.

\end{document}